\documentclass[a4paper,11pt]{article}
\usepackage{jheppub} 
\usepackage{lineno}

\usepackage{makecell}

\usepackage{xcolor}

\usepackage{array}
\usepackage{multirow}

\usepackage[all,cmtip]{xy}

\usepackage{graphicx} 
\usepackage{cancel} 
\usepackage{varwidth} 

\usepackage[normalem]{ulem}

\usepackage{tikz}
\usepackage{tikz-feynman} 
\tikzfeynmanset{compat=1.1.0}
\usepackage{amsmath} 
\usepackage{adjustbox} 

\graphicspath{{figures/}}

\title{\boldmath The phase diagram of the D1-D5 CFT and localized black holes}

\author{Ofer Aharony,}
\author{Ronny Frumkin,}
\author{and Jonathan Mehl}
\affiliation{Department of Particle Physics and Astrophysics, Weizmann Institute of Science, Rehovot
7610001, Israel}

\emailAdd{ofer.aharony@weizmann.ac.il}
\emailAdd{ronny.frumkin@weizmann.ac.il}
\emailAdd{jonathan.mehl@weizmann.ac.il}

\abstract{In this paper we analyze the phases that dominate the microcanonical ensemble at various energies in the D1-D5 CFT, which is dual to type II string theory on $AdS_3
\times S^3\times T^4$. We focus on black hole solutions, and on the dependence of the phase structure on the ratio of the size of the torus to the AdS scale; as small localized black holes (with horizon topology $S^8$) grow, they can start to fill the $S^3$ or the $T^4$ or both, and we analyze the general aspects of the transitions between the various phases of uniform and non-uniform black holes, incorporating known solutions and discussing the properties of additional unknown solutions. Some features of the transitions between these phases are similar to higher dimensional AdS spaces, while other features are different. We provide evidence that when the torus is much larger than the AdS radius, there is a large range of energies where the typical states are a novel phase, described by a lattice (in the $T^4$ directions) of black holes with horizon topology $S^5\times S^3$. In this phase the entropy is linear in the energy, with a coefficient that is of order the AdS radius.}

\begin{document}
\maketitle

\flushbottom

\section{Introduction and summary of results}

One of the general questions that can be asked about any physical theory is what a typical state at energy $E$ looks like, and what is the density of states as a function of the energy, or in other words, what is the phase structure of the microcanonical ensemble as a function of the energy. In particular one can ask this question about $d$-dimensional conformal field theories (CFTs) on $S^{d-1}$ (of radius $L$), where this question is related by the state-operator correspondence to asking about the spectrum of local operators. At large enough energies compared to the scale $1/L$ of the $S^{d-1}$, scale-invariance implies that the entropy behaves as $S(E) \propto (c_T E L)^{(d-1)/d}$ for some constant $c_T$ related to the number of degrees of freedom in the theory.
However, the behavior at lower energies is not universal.

In ``holographic CFTs'' which are dual by the AdS/CFT correspondence \cite{Maldacena:1997re,Gubser:1998bc,Witten:1998qj} to some weakly coupled and weakly curved quantum gravity theory on $AdS_{d+1}\times \mathcal{M}$, one can answer this question using classical gravity. For example, consider the $d=4$ ${\cal N}=4$ $SU(N)$ supersymmetric Yang-Mills theory in the 't Hooft large $N$ limit at large 't~Hooft coupling, dual to type IIB string theory on $AdS_5\times S^5$, where both the $AdS_5$ and the $S^5$ have a large radius of curvature $R$. As we review in section \ref{sec:microcanonical_ads5} (following the review \cite{Aharony:1999ti} and references therein), at low energies the spectrum is dominated by a gas of gravitons. As the energy is increased this is replaced by a Hagedorn gas of weakly coupled strings. When the energy is further increased the strings start interacting and eventually (when the energy goes above the 10d Planck scale) become a small black hole with $S^8$ horizon topology. As the energy is increased the horizon size becomes of order $R$, and then the solutions fill the $S^5$ and undergo a topology-changing transition to black holes with horizon topology $S^3\times S^5$ \footnote{As a matter of convention, throughout this paper we label horizon topologies so that any compact manifold wrapped by the black object is written on the right.
} that are non-uniform on the $S^5$, and this branch of solutions then connects to the branch of $AdS_5$-Schwarzschild black holes that are uniform on the $S^5$, at the point where the latter black holes have a Gregory-Laflamme-type instability \cite{Banks:1998dd, Peet:1998cr, Hubeny:2002xn, Dias:2015pda, Dias:2016eto}. As we review below, the actual transition between the dominant states in the microcanonical ensemble is a first order transition between the localized black holes with $S^8$ topology and the uniform black holes. The latter black holes dominate for all higher energies, and have the expected universal high-energy behavior. In the canonical ensemble only the graviton gas and the uniform black holes appear, with a Hawking-Page transition between them  at a temperature $T \sim 1/L$ \cite{Hawking:1982dh,Witten:1998zw}. The same theory at weak 't Hooft coupling has a similar phase diagram \cite{Sundborg:1999ue,Aharony:2003sx} but without the localized black holes, such that there is a direct transition between the Hagedorn phase and the uniform black hole phase. Other holographic theories with $d \geq 3$ have a similar phase diagram, except that the Hagedorn phase only appears when the gravitational description involves a weakly coupled string theory.

In this paper we describe the answer to this question for the $1+1$-dimensional ``D1-D5 CFT'' that is holographically dual to type II string theory on $AdS_3\times S^3\times T^4$, when it has a large central charge $c = 6 Q_1 Q_5 \gg 1$ (here $Q_1$ is the number of D1-branes and $Q_5$ the number of D5-branes, when we view the CFT as arising from the low-energy limit of $Q_5$ D5-branes wrapping $T^4$ and $Q_1$ D1-branes on top of them). When $Q_1$ and $Q_5$ are both large and mutually prime (this is the case we will focus on in this paper), there are many similarities between this case and the case discussed in the previous paragraph, but there are also some interesting differences.

Like the ${\cal N}=4$ SYM theory, the D1-D5 CFT has a conformal moduli space of exactly marginal deformations (whose structure was recently reviewed in \cite{Aharony:2024fid}), such that in one region of this parameter space the holographic description is weakly coupled and weakly curved (so it is well-approximated by type II supergravity on $AdS_3\times S^3\times T^4$), while in another region the CFT is a free orbifold (an ${\cal N}=(4,4)$ supersymmetric sigma model on $(T^{4N}/S_N)\times T^4$, where $N=Q_1 Q_5$). One minor difference from the previous case is that when $N$ has several different decompositions into a product of relatively prime integers, there are different weakly coupled and weakly curved limits of the moduli space, with different values of the string scale (and thus different properties for the Hagedorn phase). A more important difference is that one of the exactly marginal deformations of the CFT controls the size of the $T^4$ (for simplicity we will discuss only the case where the four circles have an equal radius $R_T$, the generalization to different radii is not trivial but is straightforward). From the holographic space-time point of view, this size $R_T$ can be of order the string scale, it can be larger than the string scale but smaller than the radius $R$ of $AdS_3$ and $S^3$ (which is of order $N^{1/4}$ in 6d Planck units), or (for $Q_1 \gg Q_5$) it can be larger than $R$, and as we will discuss in detail, the microcanonical phase diagram looks quite different in these different regions of the parameter space.

The phase diagram at very low and at very high energies is always similar to the $AdS_5\times S^5$ case\footnote{More precisely, this is true in the NS-NS sector of the CFT that we will be considering here. Sectors of the CFT with periodic fermions start at much higher energies, and in particular the ground states in the R-R sector correspond to zero-mass BTZ black holes.}; in the weakly coupled and weakly curved regime, at low energies the typical microcanonical states begin with a graviton gas, that transitions into a Hagedorn gas, that transitions into a small black hole;
while at very high energies we have an AdS-Schwarzschild black hole (which in this case is called the ``Bañados-Teitelboim-Zanelli'' or BTZ black hole \cite{Banados:1992wn,Banados:1992gq}). However, depending on which compact space is smaller, when the small black hole grows with the energy, it either fills the $S^3$ first or it fills the $T^4$ first. When $R_T$ is of order of the string scale, the dominant phases are always uniform on the $T^4$, and the transition in which the black hole fills the $S^3$ was recently investigated in \cite{Dias:2025csz}. As noted there, the main difference from the $AdS_5\times S^5$ case is the fact that the BTZ black hole has no Gregory-Laflamme-like instabilities, so the non-uniform black hole branch and the uniform (BTZ) black hole branch do not meet at a finite size but only at a zero-size black hole; however, in any case the meeting happens when neither phase is dominant, so this does not affect the phase diagram (that is summarized in figure \ref{fig:microcanonical_ads3_rt_sim_ls} below). When $R_T$ is larger than the string scale but smaller than $R$ there is an extra intermediate phase of black holes localized in all 10 dimensions, which then transitions (similarly to the phase diagram of black holes on $\mathbb{R}^{1,5} \times T^4$) to a phase that is uniform on the $T^4$, which then continues as in the previous case. The phase diagram for this case is summarized in figure \ref{fig:microcanonical_ads3_rt_small} below. A more interesting case is a large $T^4$, $R_T \gg R$, where the black holes fill the $S^3$ first, and there is an energy range for which they are uniform on the $S^3$ and are localized on the $T^4$ (with horizon topology $S^5\times S^3$); this case was not investigated as far as we know. We were not able to construct exact black hole solutions for this case, but we use general arguments to make conjectures about their properties and about the resulting phase diagram. 
Our main result is that there is a large range of energies (around $c (R/R_T)^4 < E L < c$) where we expect a lattice of these black holes (in the $T^4$ directions) to dominate the microcanonical ensemble, with a transition to a BTZ-dominated phase at an energy of order $c/L$. This lattice phase is characterized by an entropy that is linear in the energy, similarly to a Hagedorn phase, with the coefficient of order $L$. We do not know if this coefficient of the energy is larger or smaller than the inverse of the phase transition temperature $T = 1 / 2\pi L$; if this lattice phase has a temperature below $1/2\pi L$, it would mean that this phase is the first example in which the energy density at low energies does not satisfy the sparseness condition of \cite{Hartman:2014oaa}. The phase diagram for this case is summarized in figure \ref{fig:microcanonical_ads3_rt_large} below.

For simplicity, we discuss in this paper only solutions that carry no charges and no angular momentum; such configurations are expected to dominate the microcanonical ensemble (since adding charge or angular momentum reduces the entropy at a given energy).

We begin in section \ref{sec:conventions} with a review of the relevant properties of the ``D1-D5 CFT''. Then, 
in section \ref{review} we review the phase diagram of type IIB string theory on $AdS_5\times S^5$, and the main differences between this case and the $AdS_3\times S^3\times T^4$ case. We also review in this section the phase diagram in the free orbifold limit, which is similar to the phase diagram of the weakly coupled ${\cal N}=4$ SYM theory, with a direct transition from a Hagedorn phase with $T = 1 / (2\pi L)$ to a high-energy phase saturating the Cardy bound. In section \ref{sec:rt_small} we discuss the microcanonical phase diagram when $R_T \ll R$. We review known black hole solutions for black holes that are uniform on the $T^4$ and localized on the $S^3$ (known analytically \cite{Bah:2022pdn,Bena:2024gmp} or numerically \cite{Dias:2025csz}) and describe the resulting phase diagram in detail. Then, in section \ref{sec:rt_big} we discuss the case of $R_T \gg R$ and make conjectures about the properties of the dominant black holes in this case, and the resulting phase diagram, based on general arguments. Further support for these arguments is presented in appendix~\ref{sec:scalar_green_function}. We end in section \ref{sec:summary} with our conclusions and future directions.

\section{Conventions and parameters of the D1-D5 CFT} \label{sec:conventions}

In this section we summarize the conventions and parameter ranges used to describe the D1--D5 CFT and its gravitational dual. In most of this paper we work in the region of parameter space where the bulk is well approximated by classical type IIB supergravity, with asymptotically $AdS_{3}\times S^{3}\times T^{4}$ boundary conditions.

\subsection{Type IIB string theory on \texorpdfstring{$AdS_{3}\times S^{3} \times T^{4}$}{ads3s3t4}}

The D1-D5 CFT may be defined (in some subspace of its full conformal manifold) as the near-horizon limit \cite{Maldacena:1997re} of type IIB string theory on
$\mathbb{R}^{1,4}\times S^{1}\times T^{4}$, with $Q_{1}$ coincident D1-branes wrapped on the $S^{1}$ and
$Q_{5}$ coincident D5-branes wrapped on $S^{1}\times T^{4}$. By S-duality of type IIB string theory, this is equivalent to the near-horizon limit of $Q_1$ fundamental strings and $Q_5$ NS5-branes, which gives a background with only NS-NS fields turned on (that is useful for a worldsheet description).

The ground state of this theory (in the NS-NS sector of the CFT) is dual to type IIB string theory on
$AdS_{3}\times S^{3}\times T^{4}$, described by the following supergravity fields (in the NS--NS description)
\begin{equation} \label{eq:convension__fields_supergravity}
    \begin{split}  ds^{2}_E=l_{p\left(6\right)}^{2}\left(\sqrt{Q_{1}Q_{5}}\left(ds_{AdS_3}^{2}+ds_{S^{3}}^{2}\right)\right)+\sum_{i=1}^{4} dx_{a}^{2} ,\\
    H_{3}=2Q_{5}l_{s}^{2}\left(d{\rm vol}\left(S^{3}\right)+*_{6}d{\rm vol}\left(S^{3}\right)\right), \quad e^{\Phi}=g_{s} \ .
    \end{split}
\end{equation}
The symbol ${*_{6}}$ denotes the six-dimensional Hodge dual with respect to the
$AdS_{3}\times S^{3}$ metric. 
Since we wrote an explicit factor of $l_{p\left(6\right)}$, this metric is written in the six dimensional Einstein frame (it is easy to rewrite it in any other desired frame). The explicit factor of the string scale $l_s^{2}$ in the NS--NS three-form field strength $H_{3}$ is included to ensure the correct physical dimensions.
$Q_5$ is the quantized NS-NS 3-form flux on $S^3$, and $Q_1$ is the quantized dual 7-form flux on $S^3\times T^4$.\footnote{In the NSR formalism, string perturbation theory around this background actually computes a grand-canonical ensemble of theories with different values of $Q_1$, but this will not play any role in this paper.}

We employ global, dimensionless coordinates on a unit-size $AdS_{3}$,
\begin{equation} \label{eq:ds_ads3_dimless}
    ds_{AdS_{3}}^{2}=-\left(1+\rho^{2}\right)dt^{2}+\frac{d\rho^{2}}{1+\rho^{2}}+\rho^{2}d\phi^{2} ,
\end{equation}
where  
$\phi\in[0,2\pi)$ parametrizes the boundary circle (which may be identified with the circle that the dual CFT lives on).
We parametrize $T^4$ by dimensionful coordinates $x_{a}\in[0,2\pi R_{T})$ for $a=1,2,3,4$, taking all four circle sizes to be equal for simplicity, and without loss of generality we take $R_{T}\gtrsim l_{s}$. The complementary region
$R_{T}<l_{s}$ is related to this via T-duality. T-duality on an odd number of circles relates this to a similar background in type IIA string theory, which we will not discuss explicitly here. In addition, performing a T-duality transformation in the S-dual frame with RR background fields exchanges $Q_5$ and $Q_1$, so we can assume without loss of generality that $Q_5 \leq Q_1$ (otherwise we can shift to this dual description).

For fixed values of the quantized number of D-branes $Q_5$ and $Q_1$, the parameters appearing in \eqref{eq:convension__fields_supergravity} are not completely independent;  the six dimensional string coupling, or equivalently the relation between the
six-dimensional Planck length $l_{p(6)}$  (associated with $AdS_{3}\times S^{3}$), and the string scale, is fixed to be
\begin{equation}
l_{p(6)}^{2}=\sqrt{\frac{Q_{5}}{Q_{1}}}\,l_{s}^{2}\,.
\end{equation}
This gives a relation between the ten dimensional string coupling $g_s$ and the torus radius $R_T$,
\begin{equation} \label{eq:g_s_as_RT}
g_{s}^{2}=\frac{Q_{5}}{Q_{1}}\frac{R_{T}^{4}}{l_{s}^{4}}\,.
\end{equation}
It follows that the common curvature radius
\begin{equation}
R \equiv R_{AdS_{3}}=R_{S^{3}}
\end{equation}
satisfies
\begin{equation} \label{eq:rads_as_Q5}
R^{2}=\sqrt{Q_{1}Q_{5}}\,l_{p(6)}^{2}=Q_{5}\,l_{s}^{2}\,.
\end{equation}

Finally, in some places it will be convenient to work with \emph{dimensionful} global coordinates on $AdS_{3}$. In this convention we rescale the coordinates $t$ and $\rho$ appearing above by a factor of $R$, such that
the $AdS_3$ part of the metric takes the form
\begin{equation}  \label{eq:ds_ads3_dimfull}
    ds^{2}=-\left(1+\frac{r^{2}}{R^{2}}\right)dt^2+\frac{dr^{2}}{1+\frac{r^{2}}{R^{2}}}+r^{2}d\phi^{2},
\end{equation}
where $t$ and $r$ are dimensionful. We will use either the dimensionless coordinates \eqref{eq:ds_ads3_dimless} or the dimensionful coordinates \eqref{eq:ds_ads3_dimfull}, depending on the context.

\subsection{The parameter space}

The D1-D5 CFT has a large conformal moduli space (parameter space), which we will mention below, but in most of this paper we will focus on the one-dimensional subspace of the parameter space that is labeled by $R_T$.
We now determine the region of this subspace of the parameter space in which the classical background \eqref{eq:convension__fields_supergravity} provides a reliable description, expressed in terms of $Q_{1}$, $Q_{5}$, and the ratio $R_{T}/R$. Two independent sets of constraints must be imposed.

The first set of constraints follows from requiring the validity of the classical supergravity approximation. To neglect quantum (loop) corrections, all relevant physical length scales must be large compared to the appropriate Planck length. In particular, we impose
\begin{equation}\label{eq:constrinat_r_rt_large_lp10}
R,\,R_{T} \gg l_{p(10)}\,,
\end{equation}
where $l_{p(10)}$ is the ten-dimensional Planck length. 
Since we also consider supergravity solutions that are homogeneous on $S^{3}$, $T^{4}$, or both, one should in principle further require $R$ and $R_{T}$ to be large compared to the effective Planck lengths arising in the corresponding lower-dimensional reductions (for instance on $AdS_{3}$, $AdS_{3}\times T^{4}$, and $AdS_{3}\times S^{3}$). In the parameter range of interest, these additional constraints are automatically satisfied once \eqref{eq:constrinat_r_rt_large_lp10} holds.
When $R_{T}\sim l_{s}$, the supergravity approximation along the torus is not reliable (as discussed below). Nevertheless, classical supergravity can still be trusted for configurations that are uniform on $T^{4}$, \emph{i.e.}\ described by the $AdS_{3}\times S^{3}$ system. In this case one may relax \eqref{eq:constrinat_r_rt_large_lp10} to the weaker requirement
\begin{equation}
R \gg l_{p(6)}\,.
\end{equation}

The second set of constraints follows from requiring that stringy corrections are suppressed. When the background involves a weakly coupled string theory, even when quantum loops are negligible, massive string modes can generate higher-derivative terms in the effective action, with explicit dependence on the string length $l_s$. To avoid such $\alpha'$ corrections, we demand that when $g_s < 1$ the relevant length scales should be large compared to the string scale,
\begin{equation}\label{eq:constrinat_r_rt_large_ls}
R,\,R_T \gg l_s\,.
\end{equation}
Furthermore, when $g_s > 1$ we can perform an S-duality transformation and obtain a dual string theory with coupling $1/g_s$, whose characteristic length scale is related to the D1-brane tension, $g_s^{1/2} l_s$. Thus, for $g_s>1$ we have the requirement
\begin{equation}\label{eq:constrinat_r_rt_large_ld1brane}
R,\,R_T \gg g_s^{1/2} l_s\,.
\end{equation}
As before, if we restrict attention to configurations that are uniform on $T^4$ and described by six-dimensional supergravity on $AdS_3\times S^3$, it is sufficient to impose \eqref{eq:constrinat_r_rt_large_ls} and \eqref{eq:constrinat_r_rt_large_ld1brane} only for the scale $R$.

Combining the conditions \eqref{eq:constrinat_r_rt_large_lp10}, \eqref{eq:constrinat_r_rt_large_ls}, and \eqref{eq:constrinat_r_rt_large_ld1brane}, we obtain that the region of parameter space in which the ten-dimensional classical supergravity description is reliable is
\begin{equation} \label{eq:constiants_for_supergravity_10d}
1\ll Q_{5}\ll Q_{1}, 
\qquad 
\left(\frac{1}{Q_{5}}\right)^{1/2}\ll\frac{R_{T}}{R}\ll\left(\frac{Q_{1}}{Q_{5}}\right)^{1/4}.
\end{equation}
If instead we restrict to the six-dimensional description on $AdS_{3}\times S^{3}$ (appropriate for configurations uniform on $T^{4}$), these constraints relax to
\begin{equation} \label{eq:constiants_for_supergravity_6d}
1\ll Q_{5},
\qquad 
\frac{R_{T}}{R}\ll \left(\frac{Q_{1}}{Q_{5}}\right)^{1/4},
\end{equation}
where we restrict to the regime $R_{T}\gtrsim l_{s}$ and $Q_5 \le Q_1 $. 

\subsection{Additional constants}

We find it instructive to present the results below both in terms of bulk gravitational parameters and in terms of the boundary CFT data, in particular the central charge, rather than working exclusively with $(Q_{1},Q_{5})$. We therefore summarize here the relations among the various parameters.

The CFT central charge is related to the D-brane number by $c=6 Q_1 Q_5$. This is consistent with its relation to the three-dimensional Newton constant $G_{3}$, given up to order one numbers\footnote{We'll use $\simeq$ to denote equalities that are true up to unimportant numerical factors of order one, and $\sim$ and $\approx$ will be used for approximate functional dependence (that may also be up to some numerical factor).}, by
$\frac{1}{G_{3}}=\frac{\mathrm{Vol}(S^{3})\,\mathrm{Vol}(T^{4})}{G_{10}} \simeq \frac{R^3 R_T^4}{G_{10}}$,
through the Brown--Henneaux formula \cite{Brown:1986nw},
\begin{equation}
    c=\frac{3R}{2G_{3}}.
\end{equation}
Using \eqref{eq:rads_as_Q5} and $G_{10}\simeq l_{s}^{8}g_{s}^{2}$, the constraints can be re-expressed in terms of the dimensionless ratio $R_{T}/l_{s}$ and the CFT central charge $c$. The analog of \eqref{eq:constiants_for_supergravity_10d} becomes
\begin{equation} \label{eq:constiants_for_supergravity_10d_with_c}
1\ll Q_{5}\ll Q_{1},
\qquad
1\ll \frac{R_{T}}{l_{s}} \ll c^{1/4}\,.
\end{equation}
Likewise, the analog of \eqref{eq:constiants_for_supergravity_6d} may be written as
\begin{equation} \label{eq:constiants_for_supergravity_6d_with_c}
1\ll Q_{5},
\qquad
1 \lesssim \frac{R_{T}}{l_s} \ll c^{1/4}\,.
\end{equation}

Finally, throughout this work we often treat $R_{T}$ and $G_{10}$ (and hence the ten-dimensional Planck length $l_{p(10)}$) as independent parameters, even though they are related through \eqref{eq:g_s_as_RT}. This is consistent because one may instead specify the ratio $R_{T}/R$, and then determine the curvature scale $R$ from $c \simeq \frac{R^{4}R_{T}^{4}}{g_{s}^{2}l_{s}^{8}}$. In particular,
\begin{equation}
\frac{R}{l_{p(10)}} \simeq c^{1/8}\left(\frac{R}{R_{T}}\right)^{1/2}\,.
\end{equation}
From this viewpoint, one can tune the relative size of $l_{p(10)}$ by varying the central charge $c$, subject to the constraints in \eqref{eq:constiants_for_supergravity_10d_with_c} and \eqref{eq:constiants_for_supergravity_6d_with_c}.

Up to this point we focused only on a one-dimensional slice of the parameter space of the D1-D5 CFT; the full 84-dimensional parameter space was recently reviewed in \cite{Aharony:2024fid}. 15 additional parameters involve the shape of the torus and the NS-NS $B$-field on the torus (in the NS-NS description of the theory); as long as the torus is more or less uniform these will not affect our discussion, while for non-uniform tori the phase diagram will change, but we will not analyze it in this paper. 4 additional parameters involve turning on RR fields. By turning on these parameters, one can connect the subspace discussed above for some relatively prime values of $Q_1$ and $Q_5$, to the same subspace in the theory with some other relatively prime $Q_1'$ and $Q_5'$, as long as $Q_1' Q_5' = Q_1 Q_5$. In particular, one can connect to the theory with $Q_5'=1$ which is the free orbifold theory, a supersymmetric sigma model on $(T^4)^{Q_1 Q_5} / S_{Q_1 Q_5} \times T^4$. The remaining 64 parameters are $J{\bar J}$ deformations of the CFT, that only affect states charged under the $U(1)^{16}$ global symmetries of the D1-D5 CFT, so they will play no role in our discussion in this paper.

\section{Review of microcanonical phases in \texorpdfstring{$AdS_{5}\times S^5$}{ads5} and the differences in  \texorpdfstring{$AdS_{3}\times \mathcal{M}$}{ads3}}
\label{review}

In this section, we begin in section~\ref{sec:microcanonical_ads5} by reviewing established results concerning the phases dominating the microcanonical ensemble for string theory on $AdS_{5}\times S^{5}$. 
Then, in section~\ref{sec:difference_ads5_ads3}, we emphasize the key distinctions between this system and string theory on the $AdS_{3}\times S^3\times T^4$ background, which will be our main focus in this paper.  Finally, in section~\ref{sec:hartman_bound} we review the phase diagram of the symmetric orbifold theory.

For simplicity, from here on, we will take the radius of the sphere that the CFT lives on (a circle in the case of the D1-D5 CFT) to be equal to the radius of the bulk anti-de-Sitter space, $L=R$ (the transformation to other values of $L$ is straightforward).

\subsection{The microcanonical ensemble of \texorpdfstring{$AdS^{5}\times S^5$}{ads5}} \label{sec:microcanonical_ads5}
The Hilbert space structure of type IIB string theory on $AdS_{5}\times S^{5}$, with string coupling constant $g_s$ and string length $l_s$ (or equivalently, string mass $m_s=\frac{1}{l_s}$), in the regime corresponding to the large $N$ 't~Hooft limit where $R_{AdS_5} = R_{S^5} = R \gg l_s$ and $g_s \ll 1$, was discussed, for instance, in \cite{Banks:1998dd}, and is reviewed in section~3.4.1 of~\cite{Aharony:1999ti}\footnote{In the following, the Cauchy slice on which the Hilbert space is defined is taken to be the surface orthogonal to the timelike Killing vector $\partial_\tau$ of global $AdS_5$, in which the metric is written as
\begin{equation}
    ds^2 = R^2 \left( -\cosh^2(\rho)\, d\tau^2 + d\rho^2 + \sinh^2(\rho)\, d\Omega_3^2 \right).
\end{equation}}. 
Here, we briefly review the results for different ranges of the energy and elaborate on several relevant points. The main features are summarized in figure~\ref{fig:microcanonical_ads5}. In this case, the bulk parameters are related to those of the dual $SU(N)$ CFT by $\frac{R}{l_s}= \left(2\lambda\right)^{1/4}$ and $g_s =\frac{\lambda}{2\pi N}$, where $\lambda=g_{YM}^2N$ is the 't~Hooft coupling.

\subsubsection{Gas of free gravitons}

For low energies compared to the string scale, $ \frac{1}{R} \ll E \ll m_s$, only the lowest excitations of the strings, which correspond to ten dimensional gravitons (and their superpartners), contribute to the spectrum.   
The energy levels of a single graviton are determined by solving the linearized supergravity equations for stationary waves with a specified frequency.  
Because of the radial potential attracting energetic particles to the center, particles in $AdS_5$ (and trivially on $S^{5}$) may be viewed as effectively confined within a box of size $\sim R$. For sufficiently low energies, interactions between gravitons can be neglected, allowing us to treat the system as a free gas. Consequently, the entropy is that of a free gas of bosons and fermions (with an $O(1)$ number of species) in a box of size $\sim R$ in $9+1$ dimensions \cite{landau2013statistical},
\begin{equation} \label{eq:ads5_entropy_free_gravitons} S_{\text{gravitons}}\left(E\right)\simeq\left(RE\right)^{\frac{9}{10}} .
\end{equation}

 \subsubsection{Gas of free strings}
 
 For energies above the string mass $E>m_s$, when the string theory is weakly coupled, we have to consider also massive string excitations. In flat space, the connection between the mass of the excitation and the excitation level of free strings is given by $m=\frac{1}{l_s}\sqrt{N+A}\approx \frac{1}{l_s}\sqrt{N}$, where $N$ is the level and $A$ is a constant of order $1$. The resulting density of states is the Hagedorn density, and the entropy for $E\gg m_s$ is given by (see e.g. \cite{Polchinski:1998rq,Zwiebach:2004tj})
  \begin{equation} \label{eq:ads5_entropy_free_strings}
    S_{\text{Hagedorn}}\left(E\right)\simeq l_sE.
 \end{equation}
 In curved space-time, as long as $R\gg l_s$, we expect flat space to be a good approximation, and corrections to the flat-space Hagedorn temperature and to \eqref{eq:ads5_entropy_free_strings} can be systematically computed, and they are of order $\mathcal{O}\left(\frac{l_s}{R}\right)$ (see, for instance,  \cite{Barbon:2001di,Giveon:2005mi,Lin:2007gi,Berkooz:2007fe,Mertens:2015ola,Harmark:2021qma,Ashok:2021vww,Urbach:2022xzw,Harmark:2024ioq}). 

The transition between the previous phase and this one occurs when \eqref{eq:ads5_entropy_free_gravitons} is equal to \eqref{eq:ads5_entropy_free_strings}, namely $E \sim \frac{R^{9}}{l_{s}^{10}}$.
 Below this scale the gas of free string states is dominated by the ten dimensional free gravitons, and above it the same gas is dominated by excited massive string states (with a similar density of states arising from states containing one string or several strings).

When the string coupling is of order one (so that we are not in the 't~Hooft limit), this phase does not arise, and the graviton phase of the previous subsection transitions directly into the black hole phase descried in the next subsection.

\subsubsection{Black hole solutions} \label{sec:ad5_black_holes}

As we increase the energy further, the strings become longer, and when $g_s > 0$ their self-interactions eventually become significant, causing them to collapse into a black hole. There are several possible stationary black hole solutions that should be considered.

A configuration of strings that is localized both in $AdS_5$ and in $S^5$ is expected to collapse into a configuration resembling a ten-dimensional Schwarzschild black hole. Indeed, when the horizon radius $r_H$ obeys $l_s \ll r_H \ll R$ such a solution (with horizon topology $S^8$, and horizon radius approximately equal $r_H \sim (G_{10} E)^{1/7}$) is equal to the ten-dimensional Schwarzschild solution with small corrections, and it dominates over the other solutions that we will describe below.
Using the ten-dimensional Bekenstein-Hawking formula, its entropy scales as
 \begin{equation} \label{eq:ads5_entropy_small_BH}
    S\left(E\right)\simeq G_{10}^{1/7} E^{8/7}.
 \end{equation}
By comparing the entropies in \eqref{eq:ads5_entropy_free_strings} and \eqref{eq:ads5_entropy_small_BH}, one finds that the transition between the string phase and the ten-dimensional black hole phase occurs at $E\sim m_s/g_s^2$. This is expected to be a continuous transition, with the string gas condensing to become a black hole with a string-scale horizon $r_H \sim l_s$ \cite{Horowitz:1996nw,Susskind:1993ws}.
As the energy increases further, the horizon radius grows, and once $r_H$ becomes of order $R$, the geometry of both $AdS_5$ and $S^5$ becomes important, rendering the Schwarzschild approximation invalid. This regime was explored numerically in~\cite{Dias:2016eto}.

Another black hole solution, that exists for any energy larger than the five-dimensional Planck scale, is a five-dimensional AdS-Schwarzschild black hole times $S^5$; this solution has horizon topology $S^3\times S^5$, and is uniform over the $S^5$ compact space. When these black holes are small they resemble five dimensional black holes in flat space times $S^5$ (one can think of them as black 5-branes wrapped on $S^5$), but as their horizon radius becomes larger than $R$ their specific heat becomes positive, and these solutions are expected to dominate at very large energies.
For $r_H \gg R$, their entropy can be approximated by
\begin{equation} \label{eq:ads5_entropy_large_BH}
    S\left(E\right)\simeq G_{10}^{-1/4}R^{11/4}E^{3/4} .
\end{equation}
The exponent of the energy fits the general expectation for the high-energy behavior, as the dual CFT for this theory is of dimension $d=4$.
By comparing \eqref{eq:ads5_entropy_small_BH} and \eqref{eq:ads5_entropy_large_BH}, we expect that the transition between the phases occurs at 
 $E \sim R^7 / G_{10}$, which is also when the horizon radius of both solutions obeys $r_H \sim R$.

What does the transition between these solutions look like? As we increase the energy of the black holes with $S^8$ horizon, they gradually fill the $S^5$, and it is believed that at some critical energy the solutions undergo a topology-changing transition (similar to the black-hole/black-string transition \cite{Kol:2002xz,Wiseman:2002ti,Sorkin:2003ka,Kudoh:2004hs,Harmark:2002tr,Wiseman:2002zc,Harmark:2003dg}), through a solution with a conical singularity. After this transition the horizon topology of these black holes becomes $S^3\times S^5$, though they are not uniform on the $S^5$. Related to this, when the horizon radius of the uniform-in-$S^5$ black holes discussed in the previous paragraph obeys $r_H \ll R$, they resemble black branes wrapping an $S^5$ that is much larger than their horizon radius, so they have a Gregory-Laflamme instability \cite{Gregory:1993vy,Gregory:2011kh} towards turning on non-constant fields on the $S^5$. 
There is \cite{Hubeny:2002xn, Buchel:2015gxa,Dias:2015pda} some maximal horizon radius (energy) $r_H \approx 0.44 R$ for which this instability occurs; at this maximal radius (energy), the frequency of the non-constant fluctuations vanishes, and we can turn on this zero mode in static solutions and obtain a new family of solutions that are non-uniform on the $S^5$ \cite{Dias:2015pda}, and that continuously connect to the uniform black holes. This family of solutions is commonly referred to in the literature as `lumpy black holes'.
It is believed, and supported by numerical simulations, that the non-uniform branch that arises from the high-energy limit of the localized solutions, and the non-uniform branch that arises from the Gregory-Laflamme zero mode of the uniform black holes, are identical, such that if we continuously follow the localized solutions, they merge with the AdS-Schwarzschild solution at $r_H \approx 0.44 R$. This is analogous to what happens in the black hole-black string transition for black holes in the presence of a circle in flat space.

\begin{figure}[t]
\centering
\includegraphics[width=0.47\linewidth]{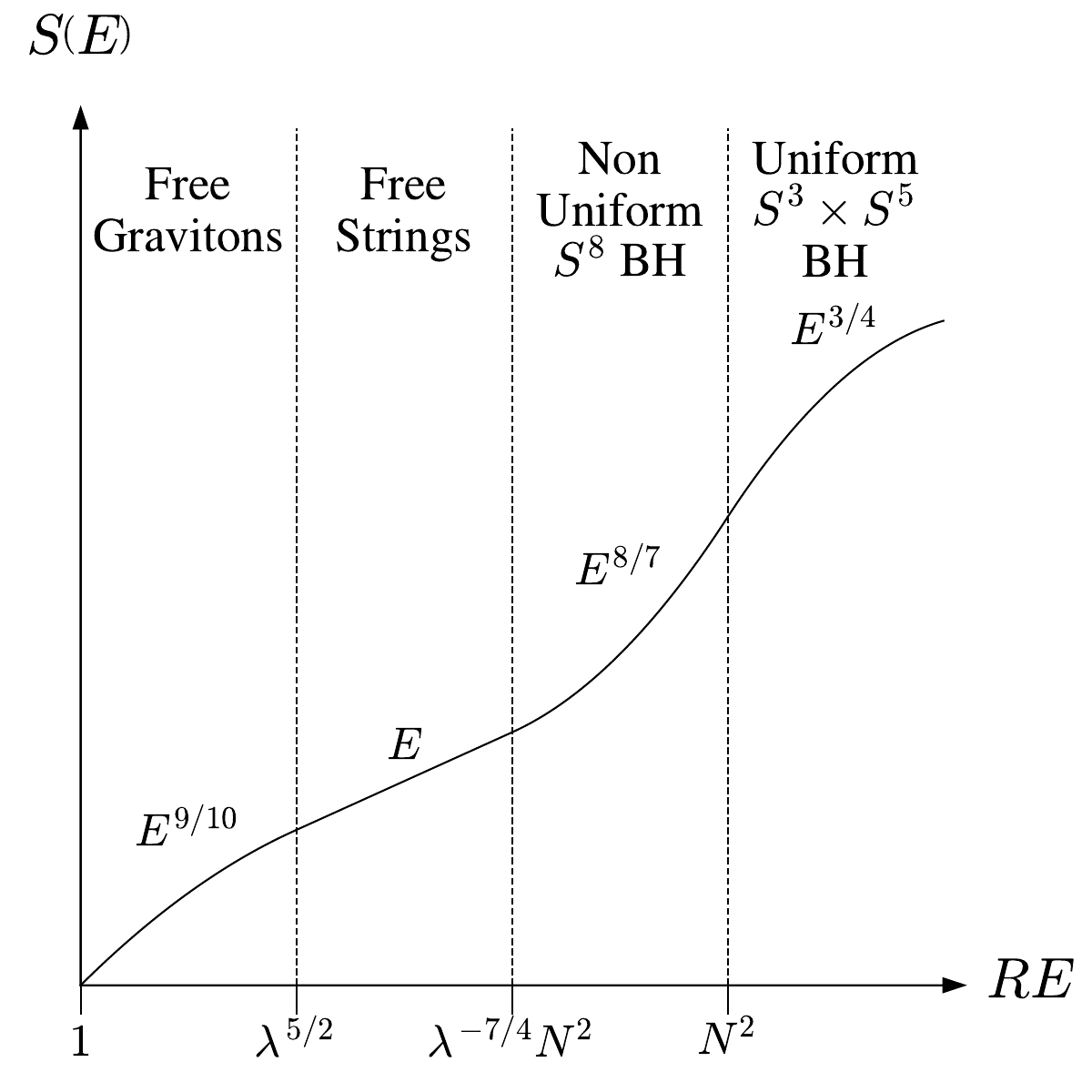}
\hspace{0.04\linewidth}
\includegraphics[width=0.47\linewidth]{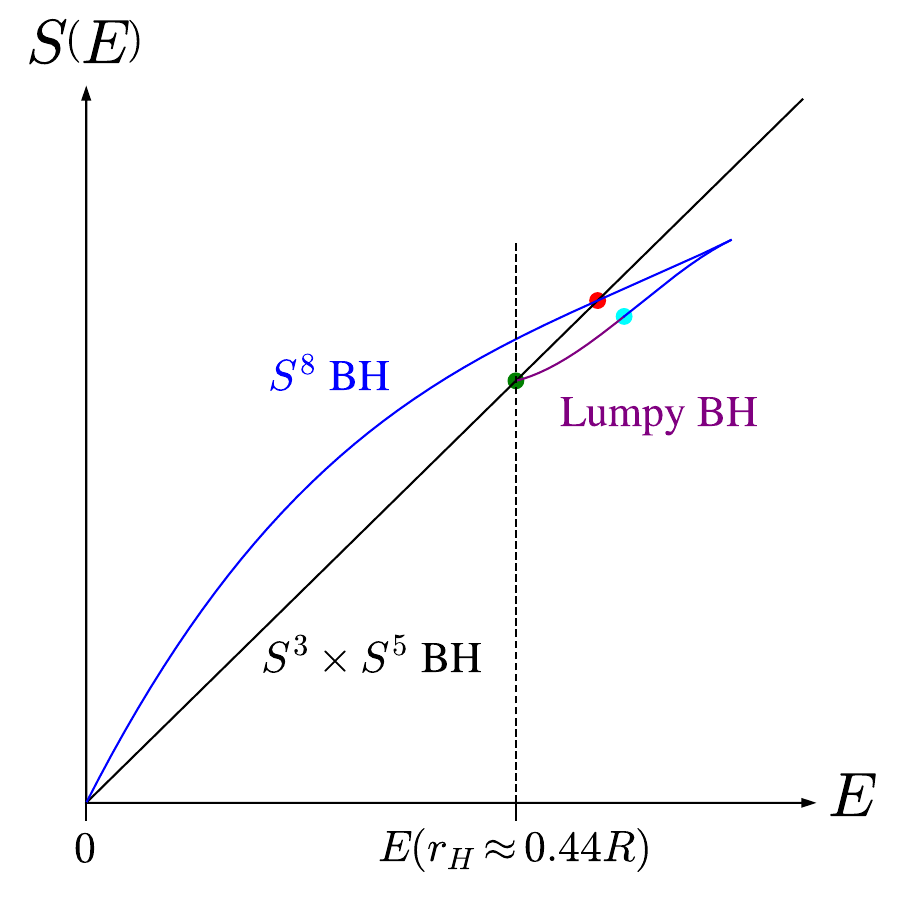}

\caption{ 
Schematic plots of the microcanonical entropy as a function of energy for string theory on $AdS_{5}\times S^{5}$ (at large $N$ and large 't~Hooft coupling $\lambda$). \textbf{Left:} The phase that dominates in each energy range. \textbf{Right:} A magnified view of the two dominant black hole phases and the transition between them. The green dot marks the location of the highest Gregory--Laflamme zero mode of the black hole that is uniform on the compact $S^{5}$, while the red dot indicates the first-order phase transition. The cyan dot denotes the conjectured topology-changing transition between the $S^{8}$ black hole branch and the lumpy black holes.
}
\label{fig:microcanonical_ads5} 
\end{figure}

In principle, this picture allows for a smooth transition between the two types of solutions, where the localized solutions that dominate at low energies merge with the uniform solution, that then dominates at higher energies. However, it turns out that the transition between the branches is actually a first order transition, at a higher energy than the energy where the solutions merge. 
This follows from the numerical results of \cite{Dias:2015pda, Dias:2016eto,Cardona:2020unx}, and we illustrate the argument in the right panel of figure~\ref{fig:microcanonical_ads5}. In particular, \cite{Dias:2016eto} shows explicitly that, at the highest  energy for which the uniform black hole exhibits a Gregory--Laflamme instability, the entropy of the localized black hole is larger. This corresponds to the blue solid curve in figure~\ref{fig:microcanonical_ads5} lying above the black curve at the point where $r_{H} \approx 0.44\,R$ (see also figure~3 of \cite{Dias:2016eto}).
As we increase the energy further, there is a point at which a uniform and a non-uniform black hole have identical energy and entropy, yet their geometries are not equivalent (see also figure 12 of \cite{Dias:2015pda}).
At this point we have a first order phase transition between the different configurations that dominate the microcanonical ensemble. 
Eventually, as we continue to follow even further, the non-uniform black hole is expected to meet the uniform branch at the Gregory-Laflamme threshold. At some point along the localized branch, we encounter the topology-changing transition where the horizon changes from $S^8$ to $S^3 \times S^5$ and we expect to get the lumpy black holes \cite{Dias:2015pda,Dias:2016eto,Cardona:2020unx}.

There are also other black hole solutions in $AdS_5\times S^5$, such as solutions that at low energies resemble a black string (a nine dimensional Schwarzschild black hole times a circle) wrapping an $S^1$ equator of the $S^5$, with horizon topology $S^7\times S^1$, or solutions related to Gregory-Laflamme instabilities with higher angular momentum \cite{Dias:2015pda}. These other solutions, like the non-uniform solutions with $S^3\times S^5$ topology mentioned above, never dominate the microcanonical ensemble. The phase diagram of the dominant phases in this ensemble is summarized in figure \ref{fig:microcanonical_ads5}. Note that in the canonical ensemble only the bottom and top phases appear, with a Hawking-Page transition between them \cite{Hawking:1982dh,Witten:1998qj,Witten:1998zw} at a temperature $T \sim 1/R$.

\subsection{Why \texorpdfstring{$AdS_3$}{ads3} is different from  \texorpdfstring{$AdS_5$}{ads5}} \label{sec:difference_ads5_ads3}

The phase diagram of string theory in other $AdS_{d}$ space-times at weak coupling, with $d>3$, is believed to be similar to the one described in the previous subsection; for string couplings of order one (or in M theory) the free string phase no longer appears, but the other phases still have a similar behavior. However, the $d=3$ case is different, and in particular
there are two key aspects in which the $AdS_3 \times S^3 \times T^4$ background that we discuss in this paper is expected to differ qualitatively from the $AdS_5 \times S^5$ system.

The first significant difference is that the size of the compact dimensions is not solely determined by the curvature scale of the AdS space. In the case of $AdS_5 \times S^5$, the negative curvature of $AdS_5$ is precisely balanced by the positive curvature of the compact space $S^5$, introducing only one additional length scale $R$ beyond the fundamental string scale. For $AdS_3 \times S^3 \times T^4$, the characteristic radii of $AdS_3$ and $S^3$ are equal, and we will focus on the case where they are large compared to the string and Planck scales, $R_{AdS_3} = R_{S^3} = R \gg l_s$. However, the size of the torus $T^4$, which we denote by $R_T$, is an independent parameter. This is most clearly visible in the case of the background with NS-NS fluxes, where the theory on the string worldsheet is just a direct product of sigma models on $AdS_3$, $S^3$ and $T^4$, and there is no constraint on the size of the $T^4$ (though very large and very small sizes are identified by T-duality)\footnote{In the S-dual description with R-R fluxes, the size of the $T^4$ in string units is fixed by the fluxes; however the ratio between its size and the size of $AdS_3$, which is independent of the duality frame, is still arbitrary. For simplicity we will focus here on the description in terms of NS-NS fluxes, and assume it is weakly coupled, but most of our results will be valid also for other configurations.}. Therefore, in the following sections, we explore several possible hierarchies among $R_T$, $R$, and $l_s$ to fully capture the parameter space of the theory.

The second aspect is that gravity in $d=3$ is different from gravity in $d>3$, because it has no dynamical gravitons and no flat-space Schwarzschild black holes. Because of this, the large black holes in the $AdS_3$ system, namely the black holes that fill uniformly the compact space $S^3\times T^4 $, are special compared to their higher dimensional counterparts in several ways. In the following paragraphs, we highlight some differences that modify the phase diagram significantly, independently of the exact details of the compact space.

One property is that the AdS$_3$-Schwarzschild black hole phase does not connect to empty AdS space when the black hole radius becomes small, as it does in higher dimensions. This can be seen directly from the BTZ metric, which describes all spherically symmetric black hole solutions in asymptotically $AdS_3$ space~\cite{Banados:1992wn}\footnote{We use an energy normalization such that the global $AdS$ solution corresponds to $E = 0$; there is another standard convention that shifts the energies by $-\frac{1}{8G_3}=-\frac{c}{12R}$ such that the extremal BTZ solution corresponds to $E=0$.}
\begin{equation} \label{eq:btz_metric_simple}
    ds^{2}=-\left(1-8G_{3}E+\frac{r^{2}}{R^{2}}\right)dt^2+\frac{dr^{2}}{1-8G_{3}E+\frac{r^{2}}{R^{2}}}+r^{2}d\phi^{2} ,
\end{equation}
where $E$ denotes the ADM energy (mass) of the black hole, and $G_{3}$ is the effective Newton constant in three dimensions. The global $AdS$ solution corresponds to $E = 0$. For $0 < 8G_{3}E < 1$, there is no event horizon; instead, the geometry exhibits a conical singularity at $r \to 0$, so it is not a legitimate solution. For $8G_{3}E > 1$, an event horizon appears, corresponding to the BTZ black holes (including the extremal case $8G_{3}E = 1$). We therefore see that there exists an energy gap between global $AdS$ and the smallest possible (extremal) black hole with $8G_{3}E = 1$. This behavior is in stark contrast to higher-dimensional $AdS$ spaces, where black holes can (classically) exist with arbitrarily small energy.

A more subtle distinction between the cases $d = 3$ and $d > 3$ involves the Gregory-Laflamme instability discussed in the previous subsection. Non-extremal BTZ black holes cannot dynamically (classically) evolve into black holes that are non-uniform in the compact dimensions via a Gregory--Laflamme instability, even though for small enough $E$ such black holes with lower energies and higher entropies do exist. Moreover, they cannot be continuously deformed into another class of (non-uniform) static black hole solutions through smooth variations of the metric. This was demonstrated in~\cite{Dias:2025csz} for black holes that are uniform along the $T^4$ (but not on $S^3$). Here we extend this conclusion to black holes that can have also nontrivial dependence on the $T^4$.

A straightforward way to demonstrate this claim is by rescaling the coordinates of the BTZ black hole in \eqref{eq:btz_metric_simple} as
\begin{equation} \label{eq:btz_coordinate_transformation}
    \bar{t}=\sqrt{8G_{3}E-1}t, \quad \bar{\phi}=\sqrt{8G_{3}E-1}\phi, \quad \bar{r}=\frac{r}{\sqrt{8G_{3}E-1}} ,
\end{equation}
to obtain the metric in the form\footnote{Note the $-1$ in the $g_{tt}$ and $g_{rr}$ components, in contrast to the $E=0$ case of \eqref{eq:btz_metric_simple}.}
\begin{equation} \label{eq:btz_after_scaling}
    ds^2 =-\left(-1+\frac{\bar{r}^{2}}{R^{2}}\right)d\bar{t}^{2}+\frac{d\bar{r}^{2}}{-1+\frac{\bar{r}^{2}}{R^{2}}}+\bar{r}^{2}d\bar{\phi}^{2} .
\end{equation}
This metric is independent of $E$, so we see that the local structure of any BTZ metric is identical\footnote{The absence of an explicit energy parameter in the local structure of the BTZ black hole is unsurprising, since any BTZ black hole can be realized as a discrete quotient of global $AdS_3$ \cite{Banados:1992gq}.}. Furthermore, in the context of metric perturbations, the relevant global structure is also very similar: the only global difference is the periodicity of the $\bar{\phi}$ coordinate, which depends on $E$ and is given by $2\pi\sqrt{8G_{3}E - 1}$. However, we do not expect the dominant non-uniform black holes along the compact dimensions to break the $SO(2)$ rotational symmetry of the $AdS_3$ factor, and also the leading instability of the solution is not expected to break this symmetry, so the corresponding perturbations that would generate such solutions should not depend on $\phi$. For solutions that are independent of $\phi$, the energy of the black hole does not enter the differential operator appearing in the linearized perturbative gravitational equations around a $BTZ \times \mathcal{M}$ background (i.e. in the equations governing Gregory--Laflamme modes), nor does it affect the determination of possible eigenvalues through the global structure (except through a trivial overall normalization factor $1/\sqrt{\int d\bar{\phi}} = 1/\sqrt{2\pi\sqrt{8G_{3}E - 1}}$ required for mode normalization). Consequently, if a solution to these equations -- representing either a Gregory--Laflamme instability or a genuine zero mode corresponding to a continuous deformation -- were to exist at some finite energy, it would necessarily persist for arbitrarily high energies, and it is therefore excluded.

The same scaling argument extends straightforwardly to the supergravity equations of motion. In type IIB supergravity, the BTZ background is obtained from \eqref{eq:convension__fields_supergravity} by replacing the global $AdS_{3}$ metric $ds^{2}_{AdS_{3}}$ with the dimensionless version of the BTZ metric in \eqref{eq:btz_metric_simple}, where the combination $G_{3}E$ is understood to be measured in units of $R^{2}$. The NS--NS three-form and the dilaton remain unchanged.
Applying the coordinate rescaling \eqref{eq:btz_coordinate_transformation} to $(t,r,\phi)$ removes the explicit energy dependence from the metric. As a result, the linearized wave equations arising from perturbations of the supergravity equations of motion are likewise energy-independent, and the arguments presented above carry over without modification.

Another argument for the absence of zero modes leading to a continuous family of static solutions connected to BTZ$\times S^3\times T^4$ is the following. The static Lorentzian solutions discussed above may easily be Wick-rotated to Euclidean time by $t = i\tau$, with the Euclidean time $\tau$ periodic for smoothness of the solution at the horizon, with a period that is the inverse of the black hole temperature. This Euclidean BTZ solution, in which the $\tau$ circle shrinks to zero in the interior, turns out to be identical to the thermal Euclidean AdS solution (the Wick rotation of the $E=0$ solution in \eqref{eq:btz_metric_simple} with $\tau$ periodically identified) in which the $\phi$ circle smoothly shrinks in the interior, by exchanging the role of the $\tau$ and $\phi$ coordinates. If we keep the periodicity of the $\phi$ coordinate fixed, this relates the Euclidean BTZ solution at a temperature $T$ to the Euclidean thermal AdS solution at a temperature $1/(2\pi R)^2 T$.
The static deformations we are considering are independent both of $t$ and of $\phi$, and depend only on $r$ and on the other compact coordinates. Thus, if such deformations existed for the BTZ solution at some energy, then similar deformations would exist for the Euclidean thermal AdS at some temperature. However, Euclidean thermal AdS is stable in supergravity, so also the Euclidean BTZ backgrounds cannot have any Gregory-Laflamme instabilities\footnote{In string theory thermal AdS becomes unstable at the Hagedorn temperature, and correspondingly also BTZ is expected to become unstable for very low temperatures inversely related to the Hagedorn temperature, where the $\phi$ circle at the horizon has a size of order the string scale, as discussed in \cite{Urbach:2023npi}.}.

As an aside, we note that the absence of the Gregory–Laflamme instability for BTZ black holes is consistent with the conjecture of Hubeny and Rangamani \cite{Hubeny:2002xn} concerning the Gregory–Laflamme instability in spacetimes of the form $B_m \times X^n$, where $B_m$ denotes an $m$-dimensional black hole and $X^n$ the $n$-dimensional transverse compact space. They conjectured that such an instability can occur only if a negative eigenvalue of the Euclidean Lichnerowicz operator on $B_m$ is equal in magnitude to the lowest eigenvalue of the Laplacian on the compact space $X^n$ (see section~3 of \cite{Hubeny:2002xn}). In three dimensions, however, gravity is non-dynamical and all modes of the Lichnerowicz operator are pure gauge; therefore, the Hubeny–Rangamani criterion cannot be satisfied, as there are no dynamical modes to compare to the modes in the compact space.

As a result of the absence of continuous deformations of the BTZ black hole, black holes that are non-uniform along the compact dimensions cannot continuously connect to BTZ black holes with finite energy, similarly to what happens in higher dimensions. The only uniform solutions they can connect to are either the global $AdS_3$ solution or the extremal BTZ configuration. This is because very small black holes that approach the $AdS_3$ phase are singular, and the extremal BTZ geometry is also singular. Consequently, the supergravity description is insufficient to capture the physics near these points.\footnote{It is worth noting that the scaling argument presented above does not apply to the global $AdS_3$ and the extremal BTZ cases, even within the gravitational description. This is most easily seen from the scaled metric \eqref{eq:btz_after_scaling}, where the $AdS_3$ and extremal BTZ geometries are the only cases in which the metric cannot be rescaled to the form of \eqref{eq:btz_after_scaling}, since the coefficients of $g_{tt}$ and $g_{rr}$ contain either $+1$ or $0$ instead of $-1$. Moreover, the extremal BTZ case is treated separately in~\cite{Banados:1992gq} from the non-extremal cases, and thus should not be directly compared to them without care.}

The conclusion of this discussion is that any continuous deformation of a black hole that preserves the $SO\left(2\right)$ and is non-uniform along the compact dimensions $X^n$ in an $AdS_3 \times X^n$ gravitational system must terminate either at the global $AdS_3$ point or at the extremal BTZ point. This behavior has been observed in~\cite{Bah:2022pdn, Bena:2024gmp, Dias:2025csz} and is illustrated in the figures throughout those works. This stands in contrast to the $AdS_5 \times S^5$ system, where continuous deformations at specific finite energies have been shown to exist~\cite{Dias:2015pda}.

\subsection{The symmetric orbifold theory} \label{sec:hartman_bound}

In this paper we mostly consider the D1-D5 CFT in the region of parameter space where the holographic description is weakly coupled and weakly curved (so it is well-approximated by type II supergravity on $AdS_3\times S^3\times T^4$, see section \ref{sec:conventions}). In this limit the CFT is not well understood in general. However, as mentioned above, this region is related through exactly marginal deformations to an ${\cal N}=(4,4)$ supersymmetric free orbifold theory $(T^{4 Q_1 Q_5} / S_{Q_1 Q_5})\times T^4$ (whose bulk dual is not well approximated by supergravity, though the theory can still be described by string theory in a hybrid formulation of the worldsheet \cite{Gaberdiel:2021njm,Eberhardt:2018ouy,Gaberdiel:2011vf,Berkovits:1999im}); the extra $T^4$ does not contribute significantly to the density of states so we will ignore it in our analysis. In the free orbifold limit the $T^{4N}/S_N$ theory still has an exactly marginal parameter $R_T/l_s$, which controls the energies of momentum and winding modes around the torus.

\begin{figure}[t]
\centering
\includegraphics[width=0.90\linewidth]{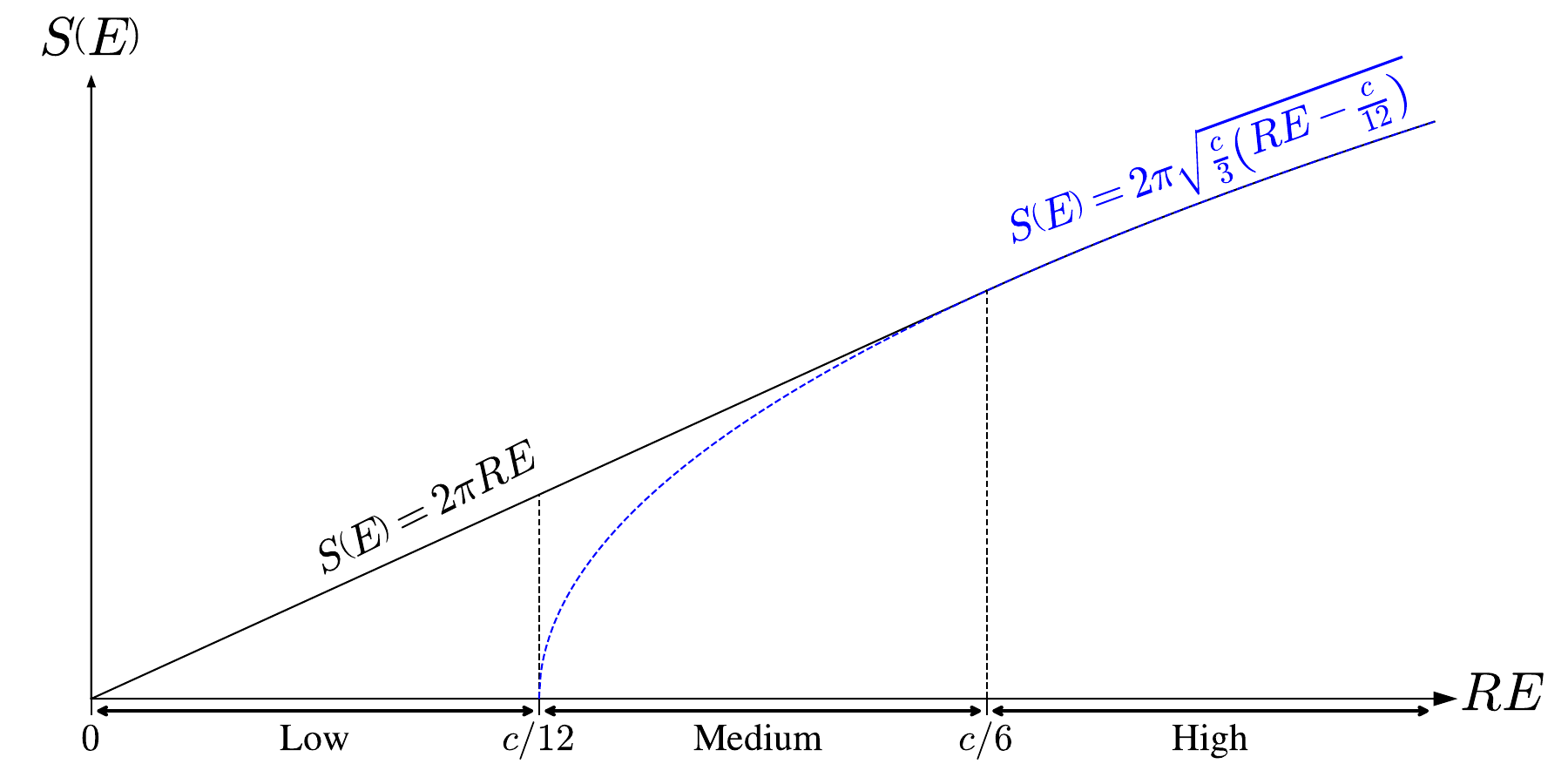}

\caption{ Entropy-Energy relation in the symmetric orbifold theories studied in \cite{Hartman:2014oaa}. The low, medium and high regimes are denoted on the horizontal axis.
}
\label{fig:cft_bound} 
\end{figure}

The phase diagram for free $\mathcal{C}^N/S_N$ orbifolds (with $\mathcal{C}$ the seed theory) in the large $N$ limit was studied in \cite{Hartman:2014oaa}.
There they classified the energy range to 3 regimes\footnote{We continue to work in the convention that the NS-NS ground state energy is $E=0$.}
\begin{itemize}
    \item Low: $0<ER<\frac{c}{12}$.
    
    \item Medium: $\frac{c}{12}<ER<\frac{c}{6}$.
    
    \item High: $\frac{c}{6}<ER$.

\end{itemize}
In both the low- and medium-energy regimes they find a linear growth of the microcanonical entropy, $S\left(E\right)\sim2\pi RE$. In the high-energy regime the entropy instead crosses over to the Cardy behavior, $S\left(E\right) \sim 2\pi\sqrt{\frac{c}{3}\left(RE-\frac{c}{12}\right)}$. The resulting phase diagram is summarized in figure \ref{fig:cft_bound}.

Furthermore, it was shown in \cite{Hartman:2014oaa} that in CFTs whose entropy in the lower range is below $S\left(E\right)\sim2\pi RE$ (this is called a ``sparseness condition''), the same expression is an upper bound on the entropy in the medium\footnote{It was also suggested that in ``generic" theories $S\left(E\right) \sim 2\pi\sqrt{\frac{c}{3}\left(RE-\frac{c}{12}\right)}$ is a lower bound in the medium energy range, but this can be violated in fine-tuned theories.} range, and the high energy range must have $S\left(E\right) \sim 2\pi\sqrt{\frac{c}{3}\left(RE-\frac{c}{12}\right)}$ for all $E \geq c/6$. The orbifold theories saturate all of these bounds. As far as we know, all theories that were studied so far obey this ``sparseness condition''; we will discuss its validity in the D1-D5 CFT in section \ref{sec:rt_big}.

\section{The microcanonical ensemble for \texorpdfstring{$R_T \ll R$}{rt smaller then rads}} \label{sec:rt_small}

In this section we analyze the phase diagram for $R_T \ll R$; we consider both the cases $l_s \sim R_T \ll R $ and $l_s \ll R_T \ll R $. The possible phases that dominate the phase diagrams are similar to those in the $AdS_5 \times S^5$ system discussed in section \ref{sec:microcanonical_ads5}, with the main differences appearing in the black hole phases.
We divide this section into three parts: the non-black hole phases, presented in section~\ref{sec:rt_small_non_bh_phases}; the dominant black hole phases, presented in section~\ref{sec:rt_small_dominent_bh}; and a discussion of general properties of black holes for this system in section~\ref{sec:rt_small_bh_discusion}.

\subsection{Low energy phases: gas of free gravitons and free strings} \label{sec:rt_small_non_bh_phases}

The dominant phase at low energies is that of a free gas of (super)gravitons. In general, the entropy of a free gas of bosons and/or fermions in $\tilde{d}$ spatial dimensions can be shown to be given by \cite{landau2013statistical}
\begin{equation}
    S \simeq V^{\frac{1}{\tilde{d}+1}}E^{\frac{\tilde{d}}{\tilde{d}+1}} , 
\end{equation}
where $V$ is the volume of the space. As in section \ref{sec:microcanonical_ads5}, any $AdS_{d+1}$ space is expected to behave effectively as a box with $d$ spatial dimensions of size $R$.

In the $R_T \sim l_s$ case, we can ignore the torus directions, as their lowest Kaluza-Klein excitations are comparable to the string scale. Thus, the system is effectively $5+1$ dimensional, and the relation between the entropy and energy for $E \gg \frac{1}{R}$ is
\begin{equation} \label{eq:entropy_gravitons_rt_small_1}
    S_{\text{gravitons without $T^4$}}\left(E\right)\simeq\left(RE\right)^{\frac{5}{6}} .
\end{equation}

For the $R_T \gg l_s $ case, there are two energy regions. For $ \frac{1}{R} \ll E \ll \frac{1}{R_T}$, the energy is not large enough to excite graviton modes with momenta in the $T^4$ directions, and the connection between entropy and energy is given by \eqref{eq:entropy_gravitons_rt_small_1}. As the energy increases to the range $\frac{1}{R_T} \ll E $, excitations in the $T^4$ directions come into play, and the entropy is given by
\begin{equation} \label{eq:entropy_gravitons_rt_small_2}
    S_{\text{gravitons  with $T^4$}}\left(E\right) \simeq\left(R_{T}^{4}R^{5}E^{9}\right)^{\frac{1}{10}}\ .
\end{equation}
A direct comparison of \eqref{eq:entropy_gravitons_rt_small_1} and \eqref{eq:entropy_gravitons_rt_small_2} shows that the latter becomes dominant when the energy reaches values of order $E \sim R^{5} / R_{T}^{6}$.

As the energy increases beyond the string scale, $E\gg m_s$, massive string excitations must be taken into account. Up to numerical factors, the result is then identical to the $AdS_{5}\times S^{5}$ case in \eqref{eq:ads5_entropy_free_strings}\footnote{There is a subtlety here for the NS-NS $AdS_3$ background, which is that the spectrum of long fundamental strings in this background is actually continuous \cite{Seiberg:1999xz,Maldacena:2000hw} above an energy $E = Q_5 / R$, and thus strictly speaking the micro-canonical density of states of this theory is infinite. In this paper we are not interested in this continuum, which describes states living near the boundary of AdS in the holographic description, and near singularities in the CFT. So, we will throw away the contributions of these long strings. This can be justified either by turning on small marginal deformations that go away from the purely NS-NS background and lift the long string continuum, or by putting some effective cutoff on the radial direction of AdS$_3$.}. Note that the properties of the string gas are approximately independent of the number of dimensions.

For $R_T\gg l_s$, one can check by comparing \eqref{eq:entropy_gravitons_rt_small_2} and \eqref{eq:ads5_entropy_free_strings} that the transition from the free-graviton gas to the free-string gas occurs at $E\sim m_s^{10}R^{5}R_T^{4}$. The same scale appears also in the $R_T\sim l_s$ case: comparing \eqref{eq:entropy_gravitons_rt_small_1} and \eqref{eq:ads5_entropy_free_strings} gives $E\sim m_s^{6}R^{5}$, and multiplying the right-hand side by $(m_s R_T)^{4}\sim 1$ yields the same parametric result.

\subsection{The dominant black hole solutions} \label{sec:rt_small_dominent_bh}

\begin{figure}[t]
\centering
\includegraphics[width=0.95\linewidth]{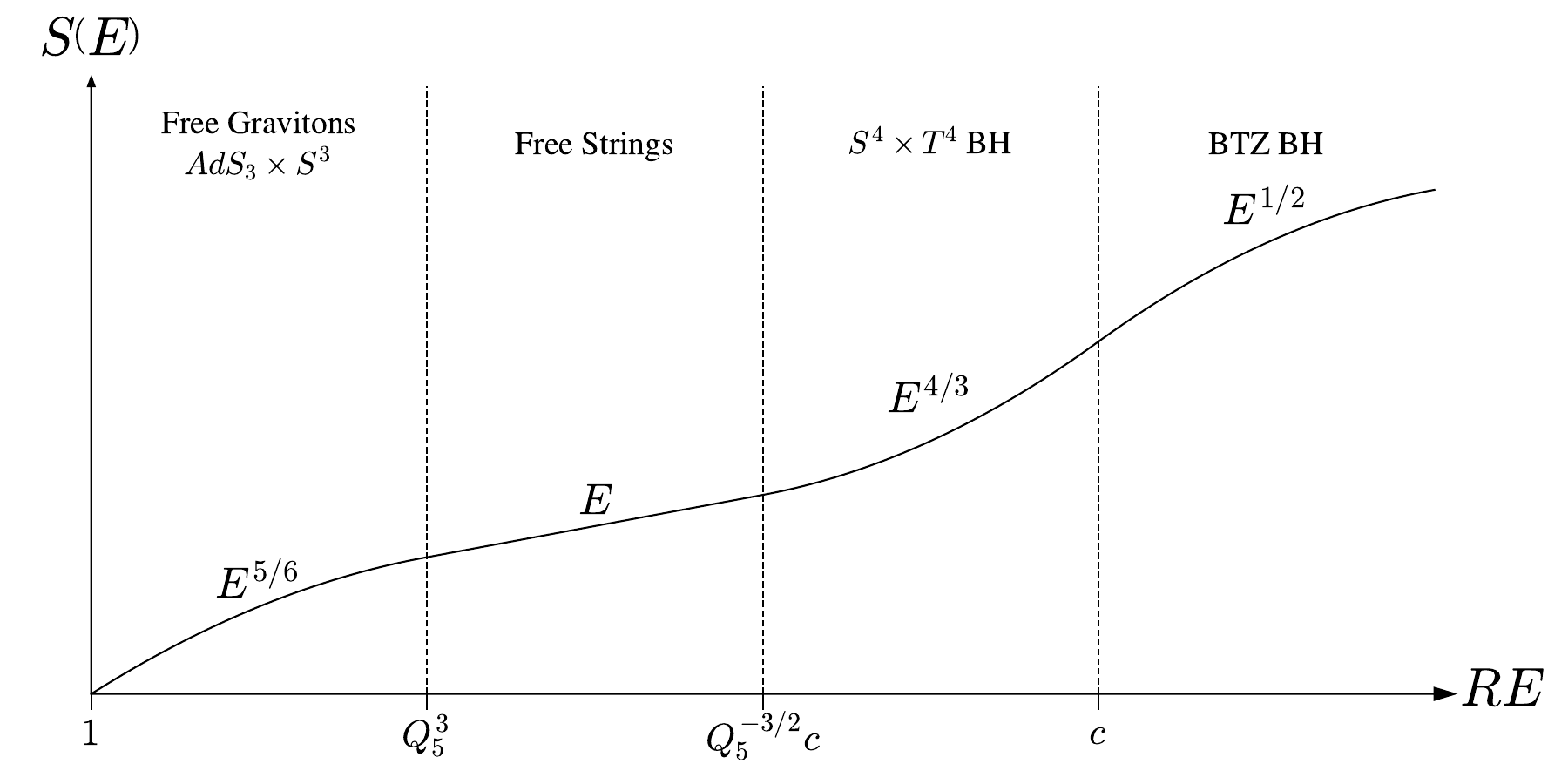}

\caption{ The entropy as a function of energy for $AdS_3 \times S^3 \times T^4$ in the microcanonical ensemble for $ R \gg R_T \sim l_s$. Note that if $Q_1$ is not large enough (compared to $Q_5$), $g_s$ is of order one (see \eqref{eq:g_s_as_RT}) and there is no phase of free strings.}
\label{fig:microcanonical_ads3_rt_sim_ls} 
\end{figure}

As in section~\ref{sec:ad5_black_holes}, once the energy increases above $E \sim m_s / g_s^{2}$, the strings begin to collapse into black holes. Once again, we expect black holes with the more `symmetric' horizon geometries to dominate the microcanonical phase space. 

We expect the following three black hole configurations to dominate, each within its own energy regime:
\begin{enumerate}
    \item
    \textbf{Black holes with horizon $S^8$}: When $R_T \gg l_s$ and the horizon radius is small compared to the other scales but larger than the string scale, namely $l_s \ll r_H \ll R_T \ll R$, we may neglect both the compactness of the $T^4$ and the curvature of the $AdS_3 \times S^3$ background. In this regime the geometry is well approximated by a ten-dimensional Schwarzschild black hole with horizon topology $S^8$, whose entropy is given by \eqref{eq:ads5_entropy_small_BH}. 

    The relation between the energy and the entropy within this approximation is
    \begin{equation}
            S(E)
            \simeq \sqrt[7]{E^{8}G_{10}}
            \simeq c\left(\frac{R_{T}}{R}\right)^{\frac{4}{7}}\left(\frac{ER}{c}\right)^{\frac{8}{7}},
    \end{equation}
    with the horizon radius $r_{H}\sim \sqrt[7]{EG_{10}}$. 
    This is valid in the range $l_s \ll r_H \ll R_T$, or equivalently for energies satisfying $m_s / g_s^{2} \ll E \ll G_{10} R_T^{7}$.

    \item \textbf{Black holes with horizon $S^4 \times T^4$}: As the energy increases and the horizon radius becomes comparable to the torus scale, $R_T \lesssim r_H \ll R$, we expect that the black holes described above that are localized on the $T^4$ cease to exist, and the black hole with a uniform mass distribution along the $T^4$ is expected to dominate the phase space. In this regime the dynamics is effectively six-dimensional,
    on the background $AdS_3 \times S^3$, and it is controlled by the six dimensional Newton constant $G_6 \sim G_{10} / R_T^4$. For radii $r_H \ll R$ one may ignore the curvature scale $R$, and the dominant black hole is well approximated by a six-dimensional Schwarzschild solution with horizon topology $S^4 \times T^4$ that breaks the $SO(4)$ symmetry of the $S^3$ to a $SO(3)$. The corresponding entropy is 
    \begin{equation} \label{eq:ads3_entropy_BH_horizen_s4t4}
        S(E) \simeq \sqrt[3]{\frac{E^4 G_{10}}{R_T^4}} \simeq c\left(\frac{ER}{c}\right)^{4/3}  ,
    \end{equation}
    with horizon radius $r_H\simeq \sqrt[3]{\frac{EG_{10}}{R_{T}^{4}}}$.
    As the horizon radius grows to $r_H\sim R$, this approximation ceases to be valid. This regime was explored numerically in \cite{Dias:2025csz}.
    
    For $R_T \approx l_s$ the string phase transitions directly into this phase. For $R_T \gg l_s$, the transition between this black hole phase and the previous one is approximately the same as the transition between ten dimensional black holes and black 4-branes wrapped on the $T^4$, related to the Gregory-Laflamme instability of the latter configurations when $R_T \gtrsim r_H$.

    \item \textbf{Black holes with horizon $S^1 \times S^3 \times T^4$}: Eventually, as the energy increases further, the horizon radius becomes large enough that the black hole fills the entire $S^3$ as well. The resulting geometry is effectively three-dimensional, described by the BTZ black hole in \eqref{eq:btz_metric_simple}. Its entropy is 
    \begin{equation} \label{eq:ads3_entropy_BH_horizen_s1s3t4}
        S(E)= \frac{2^3\pi^{4}R^{5/2}R_{T}^{2}\sqrt{E G_{10}-4\pi^{6}R^{3}R_{T}^{4}}}{G_{10}}= c\frac{\pi}{3}\sqrt{12\frac{RE}{c}-1} \, .
    \end{equation}
    This expression is valid for all $E \geq c / 12 R$ (as mentioned above, the BTZ solution starts at a non-zero energy, which in terms of the central charge is given by $ER=\frac{c}{12}$). The horizon radius is given by $r_H = R \sqrt{\frac{E G_{10}}{4\pi^{6}R^{3}R_{T}^{4}}-1}$. Comparing \eqref{eq:ads3_entropy_BH_horizen_s4t4} and \eqref{eq:ads3_entropy_BH_horizen_s1s3t4}, we see that in both cases the entropy (divided by $c$) depends only on the combination $\frac{RE}{c}$, and therefore the transition occurs at $ER \propto c \propto \frac{R^{4}R_T^{4}}{G_{10}}$. The exact numerical prefactor for the transition energy is known from the numerical analysis in \cite{Dias:2025csz}.

\end{enumerate}

\begin{figure}[t]
\centering
\includegraphics[width=0.95\linewidth]{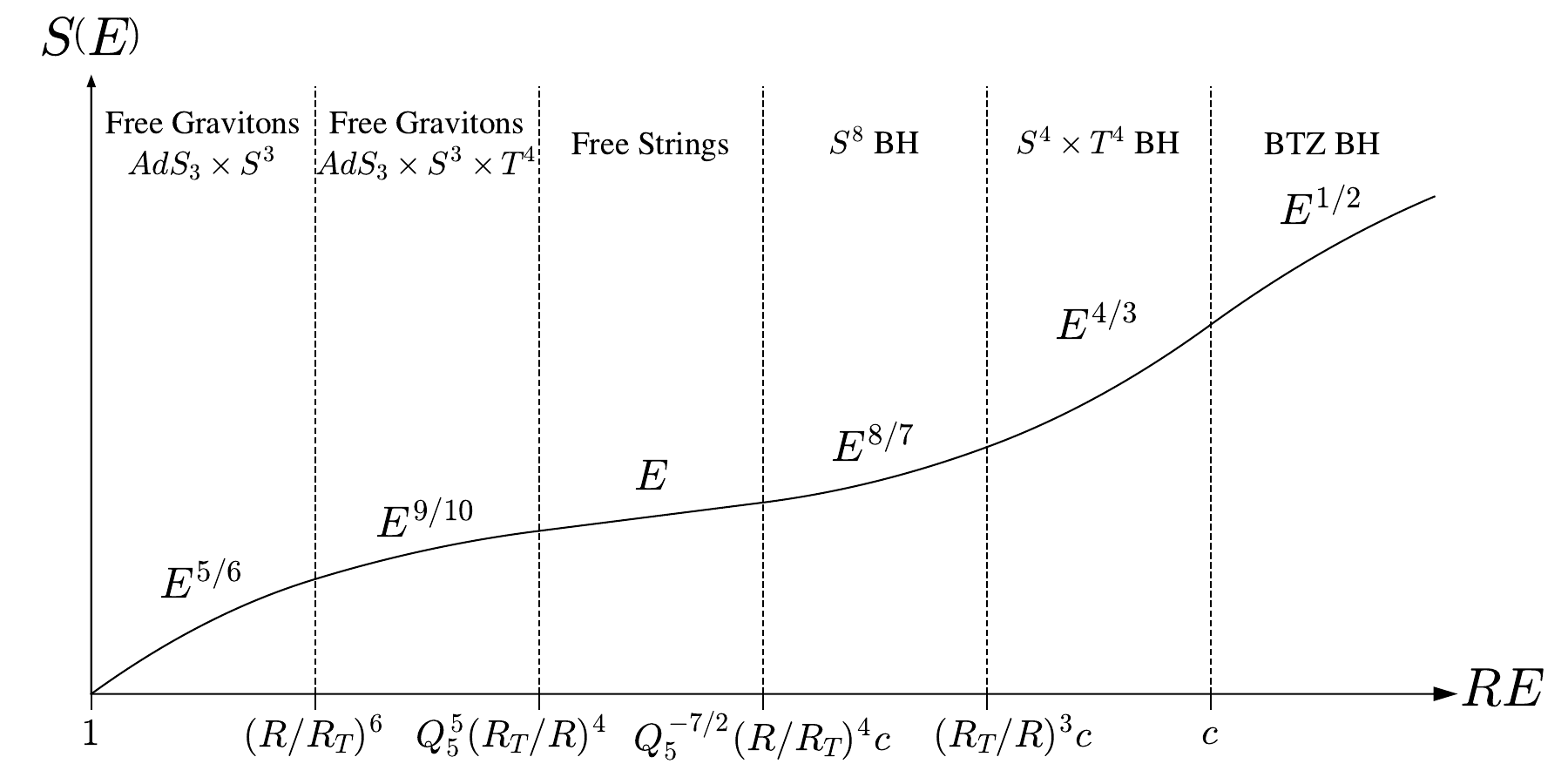}

\caption{ The entropy as a function of energy for $AdS_3 \times S^3 \times T^4$ in the microcanonical ensemble for $R \gg R_T \gg l_s$. As in figure \ref{fig:microcanonical_ads3_rt_sim_ls}, if $Q_1$ is not large enough (compared to $Q_5$), there is no phase of free strings.
}
\label{fig:microcanonical_ads3_rt_small} 
\end{figure}

We summarize the dominant phases, along with the corresponding energy ranges in which they dominate, in figures \ref{fig:microcanonical_ads3_rt_sim_ls} and \ref{fig:microcanonical_ads3_rt_small}.

\subsection{A closer look at the black hole region} \label{sec:rt_small_bh_discusion}

In this subsection we concentrate on the black hole region of the phase diagram, with particular emphasis on the features that are analogous to, or differ from, those illustrated in the right panel of figure~\ref{fig:microcanonical_ads5}.

First, we note that in addition to the black hole solutions with horizons described above, a wide variety of further configurations exists, including multi-centered solutions and black holes with more intricate horizon topologies. Although a complete classification of all solutions to the classical supergravity equations of motion is not known, several classes of such configurations have been identified in the literature.

As a simple example (in the regime $R_{T} \gg l_{s}$), one may consider black holes that wrap some, but not all, of the cycles of the $T^{4}$. Such configurations give rise to horizons with topologies such as $S^{7}\times S^{1}$, $S^{6}\times T^{2}$, and $S^{5}\times T^{3}$, and they exist only for $r_H \lesssim R_T\ll R$. At very low energies the solutions are approximately $(10-d)$ dimensional Schwarzschild black holes times $T^d$,  and their entropy can be seen to never dominate over the $S^8$-horizon solution. These solutions can, however, dominate the entropy when there is a hierarchy between the different circle sizes.

Another simple class of black holes (again in the regime $R_{T}\gg l_{s}$) consists of configurations that wrap the $S^{3}$ while remaining localized in the $AdS_{3}\times T^{4}$ directions, and whose horizon topology is $S^{5}\times S^{3}$. We elaborate on such configurations in section~\ref{sec:rt_big}.
It is evident that, in the region of parameter space considered here, these configurations never dominate. Indeed, since $R_{T}\ll R$, one may neglect curvature effects and approximate the background as $\mathbb{R}^{1,2}\times T^{4}$. At sufficiently low energies, these solutions may be viewed as Schwarzschild black holes in an effectively seven-dimensional flat space, and this approximation can be used to show explicitly that they are subdominant. Even beyond this approximation, since $r_{H}\lesssim R_{T}\ll R$, they are expected to exhibit a Gregory--Laflamme instability along the compact $S^{3}$ directions, and thus they are not favored thermodynamically or dynamically.

A large and more intricate family of black hole configurations for the $AdS_{3} \times S^{3} \times T^{4}$ system was constructed in \cite{Bah:2022pdn,Bena:2024gmp}, where complete analytic expressions are presented. The authors of \cite{Bah:2022pdn} provide a detailed recipe for constructing multi-centered black hole solutions with nontrivial topologies, some of which even mix the $T^{4}$ and $S^{3}$ directions. For example, they constructed analytic ``black pole'' solutions with horizon topology $S^3\times S^1\times T^4$, that for small energies resemble five dimensional Schwarzschild black holes wrapping an $S^1$ equator of $S^3$ (at such energies these solutions have a Gregory-Laflamme instability). These solutions break the $SO(4)$ symmetry of the $S^3$ to $U(1)\times U(1)$. Since the horizons of these new configurations are less symmetric than their counterparts in section~\ref{sec:rt_small_dominent_bh}, they are generally expected to have subleading entropies.

Second, after establishing the existence of additional black hole solutions, it should be emphasized that although the relations between entropy and energy of the dominant black hole phases presented in section~\ref{sec:rt_small_dominent_bh} are known only within specific energy ranges (see \eqref{eq:ads5_entropy_small_BH}, \eqref{eq:ads3_entropy_BH_horizen_s4t4}, and \eqref{eq:ads3_entropy_BH_horizen_s1s3t4}), we nevertheless expect black holes with the same horizon topologies to continue dominating the phase space also near the boundaries of these ranges — namely, at $E \sim G_{10}^{-1} R_T^{7}$ and $E \sim G_{10}^{-1} R_T^{4} R^{3}$. In these boundary regimes, however, the geometries can no longer be approximated by Schwarzschild or AdS–Schwarzschild solutions, and thus no simple analytic expression for their entropy is available. 
As an example, the black hole with topology $S^{4} \times T^{4}$ in the regime where $r_H$ becomes of order $R$ was studied numerically in~\cite{Dias:2025csz}, and it was found to have larger entropy than the ``black pole’’ solution, whose topology is $ S^3 \times S^1 \times T^4 $~\cite{Bah:2022pdn}.\footnote{Furthermore, \cite{Bena:2024gmp} demonstrated that uniform black holes over $T^{4}$ with more involved topologies from the family constructed in \cite{Bah:2022pdn} (or, in the terminology of \cite{Bah:2022pdn, Bena:2024gmp}, those with additional ``robes’’) have even lower entropy than the ``black pole" configuration.}

\begin{figure}[t]
\centering
\includegraphics[width=0.9\linewidth]{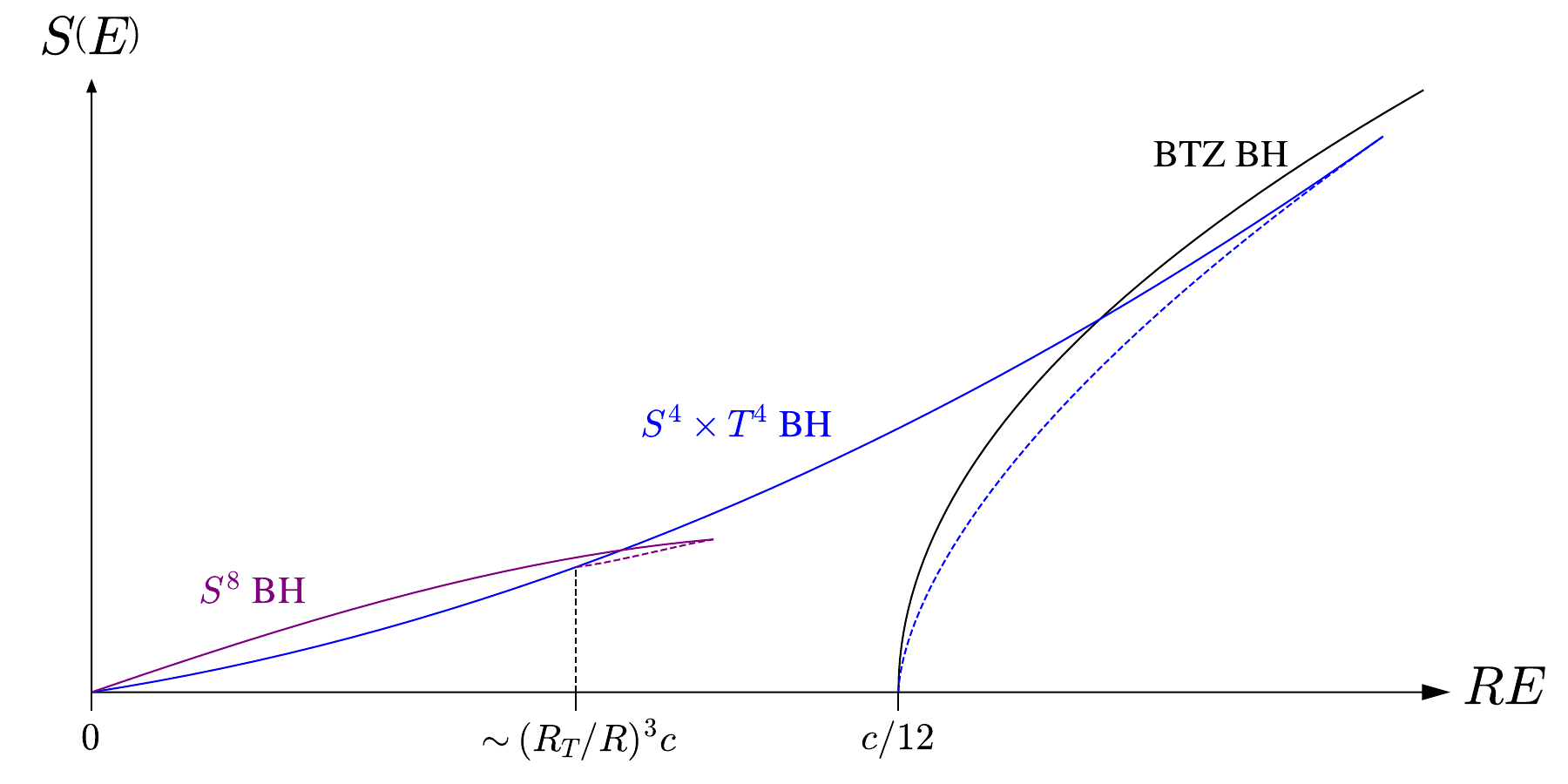}

\caption{A schematic drawing of the entropy as a function of the energy for the three types of black hole solutions that dominate the microcanonical ensemble for $R \gg R_T \gg l_s$.
}
\label{fig:ads3s3_blackholes} 
\end{figure}

Finally, we conclude this section by discussing the phase transitions between the dominant black hole branches. For $R \gg R_{T} \gg l_{s}$, the first transition occurs between the $S^{8}$ black holes and the $S^{4} \times T^{4}$ black holes. This phase transition is the same as a flat-space black-hole/black-string transition up to small corrections of order $R_T/R$, and it is expected to be qualitatively similar to the transition in the $AdS_{5} \times S^{5}$ system shown in the right panel of figure~\ref{fig:microcanonical_ads5}, as illustrated by the purple and blue curves in figure~\ref{fig:ads3s3_blackholes}.
The $S^{8}$ black hole branch begins slightly above the energy of global $AdS_{3}$. As the energy and the horizon radius increase, the black hole eventually grows to fill the $T^{4}$ manifold, causing the horizon topology to change to $S^{4} \times T^{4}$ and giving rise to non-uniform black holes over $T^{4}$, which are analogous to the ``lumpy black holes’’ studied in \cite{Dias:2015pda}.
Since the topology of these lumpy analogues coincides with that of the uniform black holes over $T^{4}$, we expect that continuous deformations of the former can end on the latter branch (represented by the blue line in the figure). 
On the other hand, since the entropy of the $S^{4}\times T^{4}$ black holes is lower than that of the $S^{8}$ black holes at sufficiently small energies, the $S^{4}\times T^{4}$ branch is expected to exhibit a Gregory--Laflamme instability throughout most of this energy range. The black-hole/black-4-brane transition in ten dimensions is of first order, as shown in \cite{Sorkin:2004qq,Kol:2006vu}, so we expect also our transition to be of first order. This in turn implies the existence of a point at which the purple and blue branches have the same entropy while corresponding to distinct geometries. Moreover, we expect that the branch of black holes that are non-uniform along $T^{4}$ (purple) joins the $S^{4}\times T^{4}$ branch (blue) precisely at the point with a Gregory--Laflamme zero mode, approaching this point from below the blue branch.

The second transition, which occurs in both $R_{T}$ regimes discussed in this section, is the transition between the $S^{4} \times T^{4}$ black holes and the BTZ black holes. This transition must be of first order. This follows from the explicit numerical results of \cite{Dias:2025csz}, which demonstrate that the $S^{4} \times T^{4}$ branch (the blue curve in figure~\ref{fig:ads3s3_blackholes}) intersects the BTZ branch (the black curve) while corresponding to a different geometry, signaling a first-order phase transition. 
Moreover, using the general arguments presented in section~\ref{sec:difference_ads5_ads3} and in \cite{Dias:2025csz}, one can see this immediately without appealing to numerics: as shown there, no continuous deformation of the BTZ metric can approach a non-extremal BTZ black hole. It seems plausible that in the region where they are sub-dominant, the localized black holes of \cite{Dias:2025csz} first undergo a topology-changing transition to fill the $S^3$ non-uniformly, and only then merge with the zero-size BTZ black hole. However, it is also possible that their topology does not change until this merger (which is the case for the analytically-known solutions of \cite{Bah:2022pdn,Bena:2024gmp}).

The same reasoning applies to the non-dominant black hole branches. These can be represented by additional curves that interpolate between global AdS and the extremal BTZ solution, as illustrated in figure~5 of \cite{Bena:2024gmp}. We do not draw them in figure \ref{fig:ads3s3_blackholes} since they are always subdominant compared to the configurations we draw.

\section{The microcanonical ensemble for \texorpdfstring{$R \ll R_{T}$}{rt larger then rads}} \label{sec:rt_big}

In this section, we analyze the phase diagram in the regime $R \ll R_T$. Many of the dominant phases are similar to those reviewed in the previous section. In particular, the BTZ-like black hole is essentially unchanged. Likewise, the free-string and $S^{8}$ black hole phases (when $r_H \ll R, R_T$) are, to leading order, insensitive to the global structure of the spacetime in the relevant regime, and their properties therefore remain the same. The $9+1$ dimensional graviton-gas phase is also unchanged, since the minimal energy of a graviton is expected to be of order $E_{\text{graviton,min}}\sim \frac{1}{R}\gg \frac{1}{R_T}$ (see section~2.2.2 of \cite{Aharony:1999ti} for the exact spectrum of scalars in $AdS$ spaces).

By contrast, the remaining black hole solutions differ and are substantially more intricate. As the localized ($S^{8}$) black hole horizon grows to the scale $r_H\sim R$, we expect it to fill the $S^{3}$ before it fills the torus, and to undergo a topology change and then a transition to a uniform-on-$S^3$ phase that is similar (but not identical) to the transitions described in the previous section. We can estimate the energy where this happens by using the ten dimensional Schwarzschild black hole solution, that gives $E R \sim c \left(\frac{R}{R_T}\right)^4$. 
A naive expectation would be that in the intermediate energy range between the regimes dominated by the $S^{8}$ black hole and by the BTZ black hole (that only starts existing at a much larger energy $E R \sim c$), the dominant phase is a black hole which is uniform along the $S^{3}$ and localized in both the $AdS$ and $T^{4}$ directions, i.e.\ with horizon topology $S^{5}\times S^{3}$.\footnote{There may also exist configurations that are localized in only some of the $T^{4}$ directions but not in others, with horizon topology $S^{1+d}\times T^{4-d}\times S^{3}$; however, they are expected to be subdominant, at least in the case we are considering here where all circles of the torus have equal sizes.
}
However, we will argue below that this is not the case, and that the phase for most of these intermediate energies is actually described by another configuration. 

In section~\ref{sec:properties_of_new_BH}, we conjecture the properties of the black holes with horizon $S^{5}\times S^{3}$ using general considerations. Concretely, we exploit the structure of the vacuum metric, together with the expected parametric scaling of the horizon radius with the energy, to argue that the entropy of a single such black hole (for high enough energies) is a concave function of the energy. Additional evidence supporting this argument is provided in appendix~\ref{sec:scalar_green_function}, where we use level sets of the static scalar Green function to gain intuition for the expected properties of the horizon geometry of such black hole solutions.

Since the entropy of a single black hole with $S^{5}\times S^{3}$ horizon is concave, it is entropically favorable for a configuration with fixed total energy $E$ to distribute the energy among several black holes. We will discuss this possibility from a thermodynamic perspective in section~\ref{sec:black_hole_lattice}, and argue that the phase diagram at these intermediate energies is dominated by a lattice of localized (on the $T^4$) black holes.

Finally, in section~\ref{sec:rt_big_summary} we present the resulting conjectured phase diagram, and comment on additional possibilities.

\subsection{Expected properties of a single black hole localized on \texorpdfstring{$T^{4}$}{T4}} \label{sec:properties_of_new_BH}

In this subsection we discuss the functional dependence of the entropy on the energy for a single black hole localized in the $AdS_{3}\times T^{4}$ directions (and uniform in the $S^3$ directions). Effectively these are black hole solutions to a 7d supergravity (the minimal $SO(3)$ gauged 7d supergravity \cite{Townsend:1983kk}), which arises as a consistent truncation of type II supergravity on $S^3$ with $H_3$ flux.\footnote{Alternatively, since we are interested in this section only in $SO(4)$-invariant classical solutions, we can consider a bosonic gravity theory that contains only the $SO(4)$-singlet fields arising in the Kaluza-Klein expansion of type II supergravity on an $S^3$ with $H_3$ flux.}

We focus on the case where the extent of the horizon in the $T^4$ directions is much smaller than $R_T$, so that the solutions are independent of $R_T$ (if we work in terms of the ten or seven dimensional Newton constant) and may be viewed as solutions in $AdS_3\times \mathbb{R}^4$. Solutions of this type involve functions of two variables, and we were not able to find them analytically\footnote{We expect that the $H_3$ flux in these solutions may differ from the vacuum solution of \eqref{eq:convension__fields_supergravity}, and that one may need also a non-constant dilaton.}; presumably they can be found numerically, using methods similar to those of \cite{Dias:2025csz}. Here we will not attempt to do this, but instead we will use general estimates of the behavior to make conjectures about the phase diagram.

First, at sufficiently small energies, the solutions and their entropy can be approximated by a seven-dimensional Schwarzschild black hole times $S^3$,
\begin{equation} \label{eq:new_black_hole_schw_approx}
    S(E)\simeq  \left(\frac{G_{10}}{R^{3}}\right)^{1/4}E^{5/4}.
\end{equation}
This approximation is valid as long as the horizon radius is smaller than the AdS curvature scale, $r_H\ll R$, which translates into the energy condition $G_{10}E\ll R^{7}$. However, in this energy range these are not the dominant solutions, and we expect black holes localized on the $S^3$, with $S^{8}$ horizons, to dominate (while the solutions we discuss have Gregory-Laflamme-like instabilities in the $S^3$ directions).

As we increase the energy, the horizon radius grows, and curvature effects in the AdS directions must be taken into account.\footnote{In a full supergravity solution, the $H$-flux that sources the curvature will also be modified, and the directions that are asymptotically flat and curved can mix in the interior of AdS. The distinction between ``flat'' and ``curved'' directions is therefore a simplifying characterization of the asymptotic geometry.} One immediate consequence of curvature is that the horizon radii along the AdS directions and along the $T^{4}$ directions will generally differ.

A simple estimate for the relation between the horizon extent along the AdS directions and along the $T^{4}$ directions follows from the vacuum metric on $AdS_3\times T^4$
\begin{equation} \label{eq:metric_global_ads_torus}
    ds^{2}=-\left(1+\frac{r^{2}}{R^{2}}\right)dt^{2}+\frac{dr^{2}}{1+\frac{r^{2}}{R^{2}}}+r^{2}d\phi^{2}+\sum_{a=1}^{4}dx_{a}^2,
\end{equation}
where $x_a \in \left[ 0, 2 \pi R_T \right]$.
The spacetime geodesic distance between two points separated in the direction of one cycle $a$ of the torus is simply $|x_{2}-x_{1}|$. By contrast, for two points separated along the AdS radial direction with $|r_{2}-r_{1}|\gg R$, the geodesic distance is
\begin{equation}
\int_{r_{1}}^{r_{2}}\frac{dr}{\sqrt{1+r^{2}/R^{2}}}\simeq R\log \left(\frac{r_{2}}{R}\right)-R\log\left(\frac{r_{1}}{R}\right) .
\end{equation}
We therefore expect the maximal extents of the horizon in the $r$ and $x$ directions (for a black hole centered at $r=x_a=0$), that we will denote by $r_H$ and $x_H$, to be related when $r_H \gg R$ as
\begin{equation}\label{eq:new_BH_x_as_r_max}
    x_{H}\sim R\log\!\left(\frac{r_{H}}{R}\right).
\end{equation}

Equation \eqref{eq:new_BH_x_as_r_max} has a powerful consequence. Since the choice of coordinate system for the AdS part of \eqref{eq:metric_global_ads_torus} (for instance, we can choose the gauge in which $r$ is defined by the $g_{\phi \phi}$ component of the metric) matches that of the BTZ metric \eqref{eq:btz_metric_simple}, we expect the relation between $r_{H}$ and the black hole energy to be qualitatively similar to the BTZ case, for which $r_{H}\sim R\sqrt{\frac{RE}{c}}$. At the very least, we expect $r_{H}$ to scale with the energy in a power-law fashion rather than exponentially. Under this assumption, together with \eqref{eq:new_BH_x_as_r_max}, it follows that $x_{H}$ depends only logarithmically on the energy. Consequently, up to energies that are exponentially large in $R_T/R$, the black hole size remains much smaller than the compactification scale, and the full metric near the black hole horizon is expected to depend only weakly on $R_T$. In other words, the black hole solution is effectively insensitive to the compactification scale over this very wide energy range.

We can now use this observation to conjecture the relation between energy and entropy. The entropy is dimensionless and scales as the horizon area divided by $G_{10}$. The horizon area is determined by the metric, and the energy enters the metric only through the combination $G_{10}E$, as is evident from the Einstein equations\footnote{Conversely, the classical solution with a specific horizon size is independent of $G_{10}$, while the energy of a given solution is proportional to $1/G_{10}$.}. Dimensional analysis therefore implies that the entropy can be written as
\begin{equation} \label{eq: singlr black hole general form}
    S=\frac{R^{8}}{G_{10}}f\left(\frac{G_{10}E}{R^{7}}\right) \simeq c\frac{R^{4}}{R_{T}^{4}}f\left(\frac{ER}{c}\cdot\frac{R_{T}^{4}}{R^{4}}\right)
\end{equation}
for some dimensionless function $f$.

We expect that once the black holes are larger than $R$, namely $x_H, r_H \gg R$, the function $f$ should have a simple behavior such as $f(\mu) \sim \mu^a$ (for $\mu \gg 1$, up to multiplicative logarithmic corrections)\footnote{This should hold up to energies that are exponential in $R_T/R$, beyond which the compactness of $T^{4}$ is expected to become important.}. Using this we obtain
\begin{equation} \label{eq:new_bh_entropy_as_a}
S\sim c\left(\frac{R}{R_{T}}\right)^{4\left(1-a\right)}\left(\frac{ER}{c}\right)^{a} .
\end{equation}
The exponent $a$ in \eqref{eq:new_bh_entropy_as_a} cannot be arbitrary. At sufficiently high energies the Cardy bound $S\left(E\right) \le \sqrt{\frac{4\pi^2}{3}cRE}$  must eventually apply, and we find that $a\le \frac{1}{2}$. Note that we do not assume that the Cardy behavior sets in already at $ER\ge c/6$, as in \cite{Hartman:2014oaa}. Instead, we only need to assume that the Cardy bound holds above some sufficiently large energy scale, which may depend polynomially (but not exponentially) on $R_{T}$.

The constraint $a\le \frac{1}{2}$ has two important ramifications. First, it implies that the entropy--energy relation for a single localized black hole is concave for energies of order $E\sim \frac{c}{R}$. Second, when $R_T\gg R$, the prefactor in \eqref{eq:new_bh_entropy_as_a} indicates that this phase is subdominant at energies of order $E\sim \frac{c}{R}$ (compared to the uniform-on-$T^4$ solutions discussed in the previous section, that have $S(E) \sim c$ for such energies, and at least one of them exists for any such energy).

The general assumptions above are sufficient to derive most of the conclusions presented in the next section. Nevertheless, we can also formulate a stronger conjecture. Since, as indicated by \eqref{eq:new_BH_x_as_r_max}, the black hole spreads primarily along the AdS directions, it is natural to expect it to share also quantitative features with black holes in $AdS_{3}$, namely BTZ, up to logarithmic corrections associated with the $T^{4}$ directions. In particular, we conjecture that the relation between the horizon radius and the energy takes the form
\begin{equation} \label{eq:new_BH_area_r_horizen_as_energy_no_dimensions}
    r_{H} \sim 
    R\sqrt{\frac{EG_{10}}{R^{7}}}\ \log^{\beta_4}\left(\frac{EG_{10}}{R^{7}}\right) \sim
    R\sqrt{\frac{ER}{c}\left(\frac{R_T}{R}\right)^{4}}\ \log^{\beta_4}\left(\frac{ER}{c}\left(\frac{R_T}{R}\right)^{4}\right),
\end{equation}
and that the relation between the entropy and energy is
\begin{equation}  \label{eq:new_BH_entropy_energy_general_arguments}
    S\sim\sqrt{\frac{ER^{9}}{G_{10}}}\log^{\alpha_{4}}\left(\frac{EG_{10}}{R^{7}}\right)\sim c\sqrt{\frac{ER}{c}\left(\frac{R}{R_{T}}\right)^{4}}\log^{\alpha_{4}}\left(\frac{ER}{c}\left(\frac{R_T}{R}\right)^{4}\right) ,
\end{equation}
for some constants $\alpha_4$ and $\beta_4$. This argument is heuristic and it is not clear if it is reliable; in appendix~\ref{sec:scalar_green_function} we discuss a more detailed (though still heuristic) estimate for the shape of the horizon, using level sets of the scalar Green function on $AdS_{3}\times \mathbb{R}^{4}$, and we find that it leads to similar relations (for specific values of $\alpha_4$ and $\beta_4$).

As a final comment, note that the arguments above are expected to hold up to energies that are exponential in $R_T/R$, namely $E \sim \frac{R^{7}}{G_{10}} e^{R_T/R}$, when the size of the black hole in the $T^4$ will become of order $R_T$ so that it will feel the compactness of the torus. At such energies, the logarithmic factor in \eqref{eq:new_BH_entropy_energy_general_arguments} behaves as $\log^{\alpha_{4}}\!\left(\frac{ER}{c}\left(\frac{R_T}{R}\right)^{4}\right)\sim \left(\frac{R_T}{R}\right)^{\alpha_{4}}$, and the entropy scales as $S\sim \left(\frac{R_T}{R}\right)^{\alpha_{4}-2}\left(cER\right)^{1/2}$. Since for such high energies we expect the Cardy bound (which is saturated by BTZ black holes) to apply, consistency requires $\alpha_4<2$. Indeed, the results obtained in appendix~\ref{sec:scalar_green_function} are consistent with this criterion.

\subsection{Lattice of localized black holes on \texorpdfstring{$T^{4}$}{T4}} \label{sec:black_hole_lattice}

As discussed above, we expect the entropy--energy relation for black holes localized on $T^{4}$ in the regime $r_H \gg R$ to be concave (and in particular this is true for the heuristic estimate \eqref{eq:new_BH_entropy_energy_general_arguments}). By contrast, at low energies with $r_H \ll R$, the entropy-energy relation is given by \eqref{eq:new_black_hole_schw_approx}, and it is convex. It follows that there exists an energy $E_{*}$ and entropy $S_{*}=S(E_{*})$ that maximize the ratio $S(E)/E$.

Therefore, on purely thermodynamic grounds (and neglecting gravitational interactions between black holes that are separated in the $T^4$ directions), for a fixed total energy $E \gg E_*$, the configuration that maximizes the entropy consists of $n= E/E_{*}$ identical black holes. The resulting entropy is then linear in both $n$ and $E$, namely $S(E)=S_{*}\,E/E_{*}$.\footnote{The possibility that multi-center black configurations can dominate over a single-center object has been discussed extensively in the literature; see for example
\cite{Gauntlett:2004wh,Denef:2007vg,deBoer:2008fk}. A related analysis in the $AdS_{3}\times S^{3}$ context appears in \cite{Bena:2011zw}.}

We conclude that, insofar as gravitational interactions can be neglected, the leading entropy at energies $E\gg E_{*}$ is expected to scale linearly with $E$. Note that this conclusion does not rely on the explicit form of \eqref{eq:new_BH_entropy_energy_general_arguments}, but only on the concavity of $S(E)$ in the relevant regime.

We refer to this phase as a ``lattice'' of black holes localized on $T^{4}$, uniform on $S^3$, and we expect the centers of all constituent black holes to lie at or close to $r=0$ in $AdS$. Black hole lattices that fill up flat space were considered in the literature (see \cite{Bentivegna:2018koh} and references therein), but these are expected to be unstable, both dynamically and thermodynamically. Our construction, on the other hand, fills up only the compact directions and is localized on a non-compact space, similar to the construction in \cite{Harmark:2003yz,Dias:2007hg}.\footnote{It is somewhat similar to localized black holes on $ \mathbb{R}^{1,3}\times S^1$ \cite{Kol:2002xz,Wiseman:2002ti,Sorkin:2003ka,Kudoh:2004hs,Harmark:2002tr,Wiseman:2002zc,Harmark:2003dg}. In these cases, a single black hole on the $S^{1}$ is constructed. By imposing periodic boundary conditions, it is straightforward to generalize such solutions to an arbitrary number $N$ of black holes. However, on $\mathbb{R}^{1,3}\times S^{1}$, we expect such configurations to be unstable.
} Since two such black holes cannot classically merge (since it would give a configuration with lower entropy), we expect that the black holes repel each other (at least at distances of order the horizon scale), and thus we expect the minimal energy configuration to be an evenly-spaced lattice.

Using the notation of \eqref{eq: singlr black hole general form}, we expect the energy of an individual black hole in the lattice to be characterized by $\mu_*=\frac{E_*R}{c}\cdot\frac{R_{T}^{4}}{R^{4}}$, where $\mu_*$ maximizes the ratio $\frac{f(\mu)}{\mu}$. Since $f(\mu)$ is dimensionless, we expect both $\mu_*$ and $f(\mu_*)$ to be of order unity, meaning in particular that $E_* R \simeq c R^4 / R_T^4$ (note that, as expected, $c/R_T^4$ depends on $R$ and $G_{10}$ but is independent on $R_T$, as discussed in section \ref{sec:conventions}). The entropy is then given by
\begin{equation} \label{eq:entropy_of_lattice}
    S\left(E\right)=c\frac{E}{E_*}\frac{R^{4}}{R_{T}^{4}}f\left(\mu_*\right)=c\frac{ER}{c}\frac{f\left(\mu_*\right)}{\mu_*} = E R \frac{f\left(\mu_*\right)}{\mu_*},
\end{equation}
and we expect this to be valid as long as $E \gg E_*$ such that the number of black holes is large, and as long as the distance between the black holes is large enough to neglect their interactions (we will analyze this below). This gives a Hagedorn-like behavior, with an effective temperature of order $1/R$. We do not know whether the coefficient of the energy in \eqref{eq:entropy_of_lattice} is smaller than $2\pi R$, in which case this configuration obeys the sparseness condition of \cite{Hartman:2014oaa}, or if it is larger, in which case this would be the first example that we know of that does not obey this condition.

We can now use \eqref{eq:entropy_of_lattice} to argue that there is a large range of energies where the proposed lattice configuration dominates over the black hole that is uniform on $T^{4}$ and localized on $S^{3}$, with horizon topology $S^{4}\times T^{4}$ (see \eqref{eq:ads3_entropy_BH_horizen_s4t4} and the discussion around that equation).  
For energies of order $E \sim \frac{c}{R}$, the black hole that is uniform on $T^{4}$ and localized on $S^{3}$ has an entropy of order $S\sim c$. Equation \eqref{eq:entropy_of_lattice} implies that the lattice entropy at these energies is also of the same order of magnitude, up to the prefactor $\frac{f(\mu_*)}{\mu_*}$.
Since we do not know its exact value, we cannot determine whether the lattice entropy is larger or smaller than that of the $S^{4}\times T^{4}$ black hole configuration at these energies. However, the lattice entropy grows linearly with the energy, whereas the entropy of the black hole with horizon topology $S^{4}\times T^{4}$ is convex for energies below $\sim c/R$ (due to the $E^{4/3}$ scaling in \eqref{eq:ads3_entropy_BH_horizen_s4t4}, and the explicit numerical results of \cite{Dias:2025csz}). It therefore follows that there is a large range of energies $c R^4/R_T^4 \lesssim E R \lesssim c$ for which
the lattice phase has larger entropy, and we expect it to be the dominant configuration in the microcanonical ensemble.

The linear dependence of the lattice entropy on the energy, and the validity of \eqref{eq:entropy_of_lattice}, are limited by the condition that the typical separation between black holes along $T^{4}$ must be larger than the size of each individual black hole. This condition can be checked using the relation \eqref{eq:new_BH_area_r_horizen_as_energy_no_dimensions} with the explicit conjectured relation \eqref{eq:new_BH_x_as_r_max}. The horizon extent of each black hole in the $T^{4}$ directions is $x_H\sim R\log(\mu_*)$, while the spacing between the black holes is given by
\begin{equation}
    \left(\frac{{\rm Vol}\left(T^{4}\right)}{n}\right)^{1/4}=
    \left(\frac{\left(2\pi R_{T}\right)^{4}}{\frac{E}{E_*}}\right)^{1/4}=2\pi
    R\left(\mu_*\frac{c}{ER}\right)^{1/4},
\end{equation}
and so for the black holes to be far enough, the energy must obey
\begin{equation} \label{eq:mu_constraint}
    \mu_* \gg \frac{ER}{c} \  \ln^{4}(\mu_*) .
\end{equation}
So we see that the lattice phase is reliable when $ER \ll c$, but   starts getting large corrections for $ER \sim c$ and we cannot control it there. However we can still argue that it dominates over a large range of energies below this, as mentioned above.

We do not know what happens once \eqref{eq:mu_constraint} is no longer satisfied. A plausible option is that the black hole horizons merge together by one or more topology-changing transitions, that eventually lead to a phase with $S^1\times S^3\times T^4$ topology that is non-uniform on the $T^4$, and that this phase eventually merges with the BTZ black hole at its minimal energy, similarly to what happened to the localized black holes discussed in the previous section.

However, what we can do is give a lower bound on the entropy coming from localized black holes at these and higher energies, by computing
the entropy of lattices that contain black holes with individual energies that correspond to $\mu>\mu_*$ and that are still far enough from each other so that their interactions can be neglected. This computation is outlined in appendix \ref{sec:non_optimal_lattice}.

\subsection{The conjectured phase diagram and other possibilities} \label{sec:rt_big_summary}

\begin{figure}[t]
\centering
\includegraphics[width=0.95\linewidth]{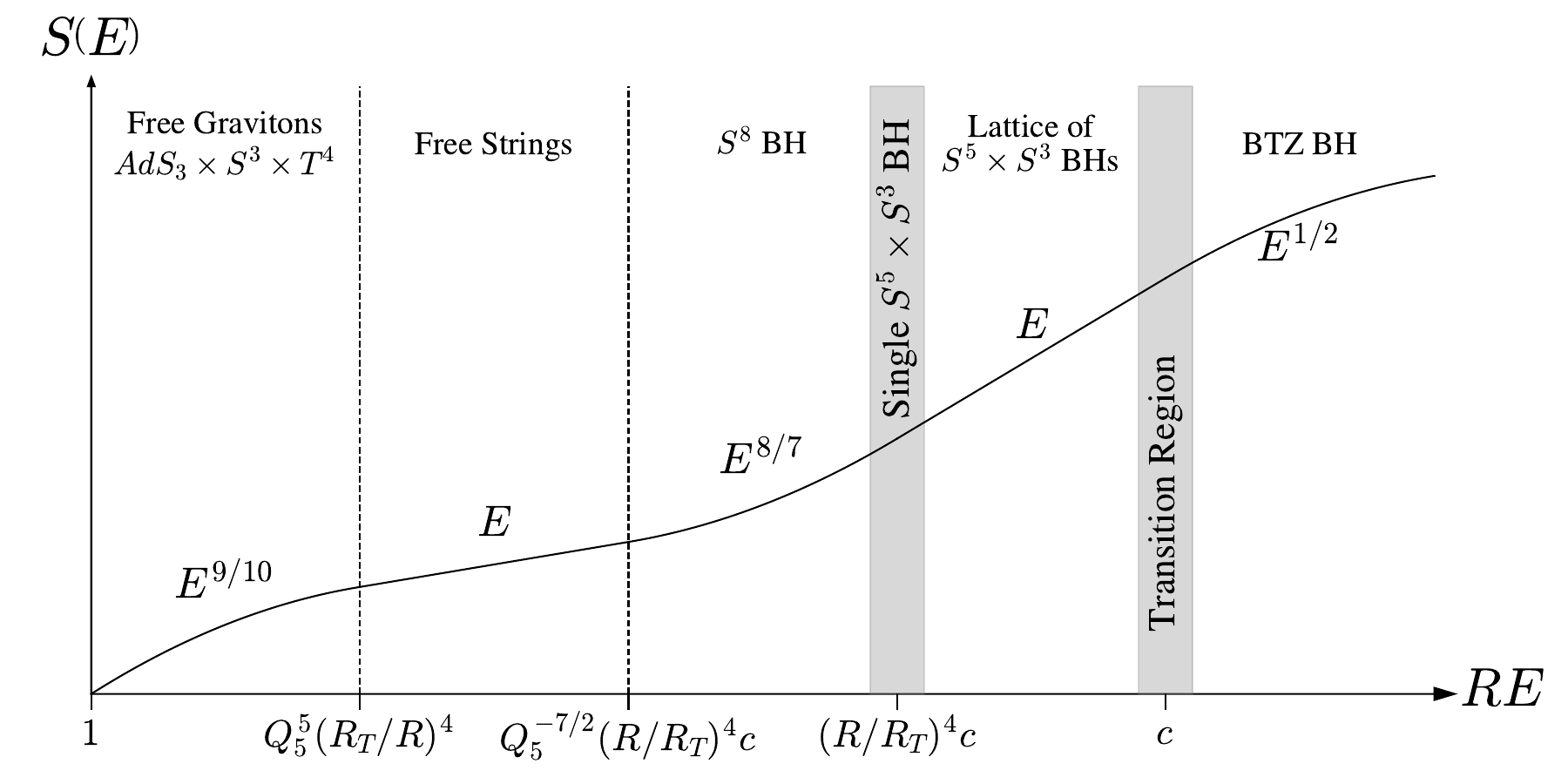}

\caption{The entropy as a function of energy for $AdS_3 \times S^3 \times T^4$ in the microcanonical ensemble, for the case $R_T\gg R$. While the other phases are well established, the lattice phase and its connection to the other phases are conjectured. As in figure \ref{fig:microcanonical_ads3_rt_sim_ls}, if $Q_1$ is not large enough (compared to $Q_5$), there is no phase of free strings, and this is also required to be able to trust ten dimensional supergravity when $R_T \gg R$.
}
\label{fig:microcanonical_ads3_rt_large} 
\end{figure}

We can now use the results obtained in this section to sketch the conjectured microcanonical phase diagram for the regime $R_T \gg R$. The result is shown in figure~\ref{fig:microcanonical_ads3_rt_large}. Relative to the $R_T \ll R$ case (figure~\ref{fig:microcanonical_ads3_rt_small}), the main difference is the intermediate phase between the $S^{8}$ black hole and the BTZ black hole: rather than a black hole that is uniform on $T^{4}$ with horizon topology $S^{4}\times T^{4}$, we find the possibility of a lattice (or, at sufficiently low energies, a single representative) of black holes that are uniform on $S^{3}$ and localized along the $T^{4}$ directions. Aside from the lattice phase, the structure of the diagram is essentially what one might naively expect.

Note that our lack of control over the solutions with $E R \sim c$ leaves open the possibility of additional dominating phases at energies of this order (before the BTZ black hole starts dominating), which could either be a lattice with fewer black holes localized on the $T^4$, or black holes that fill the $T^4$ but non-uniformly, or the black holes that are uniform on the $T^4$ but not on the $S^3$.

Although we regard the conjectured phase diagram in figure~\ref{fig:microcanonical_ads3_rt_large} as the most reasonable proposal for the dominant phases in the microcanonical ensemble, there are several potential concerns with the derivation above that could modify the final picture.

One set of concerns involves the lattice phase. In deriving the lattice entropy, we neglected the gravitational potential energy between the black holes. We expect forces between such black holes to decay exponentially with their distance on the $T^4$. However, possibly these interactions are not negligible, so that the relation between the total energy of a lattice configuration and the energy of an individual constituent differs significantly from what was assumed in section~\ref{sec:black_hole_lattice}. In that case, the entropy--energy relation could deviate from the linear behavior we predicted.
Another possibility is that the lattice phase is not dynamically stable\footnote{The construction in \cite{Harmark:2003yz,Dias:2007hg}, where the non-compact dimensions are flat, is not stable. However, we believe that by using $AdS$ asymptotics instead of flat, one can stabilize such a configuration.}. We know that the black holes must repel each other at short distances for consistency, but this could perhaps change at large distances. If the black holes attract one another, the lattice configuration would not be expected to be stable and, consequently, would also not be thermodynamically favored.

Nevertheless, even if the lattice picture is incorrect, it is plausible that some such additional phase, perhaps one with a more intricate topology, dominates the microcanonical ensemble over at least part of the energy range that we have associated with the lattice.
One reason is because, in the absence of such a phase, we obtain a contradiction. The black hole discussed in the previous section, that is uniform on $T^{4}$ and localized on $S^{3}$, dominates over the single black hole localized on $T^4$ for energies above (comparing \eqref{eq:ads3_entropy_BH_horizen_s4t4} with \eqref{eq:new_BH_entropy_energy_general_arguments} and ignoring logarithmic factors)
\begin{equation}
    E R \gtrsim \left(\frac{R}{R_{T}}\right)^{12/5}\frac{R_{T}^{4}R^{4}}{G_{10}}
    \simeq c \left(\frac{R}{R_{T}}\right)^{12/5}.
\end{equation}
However, this is problematic. At such energies, the horizon radius of the black hole that is uniform on $T^{4}$ is $r_{H}\simeq \left(\frac{R}{R_{T}}\right)^{4/5}R \ll R$. In this regime, the solution can be approximated as a black hole in $\mathbb{R}^{1,5}\times T^{4}$ that is smeared on the $T^4$, and it is therefore susceptible to the usual Gregory--Laflamme instability along the $T^{4}$ directions. As a result, it is expected to be subdominant relative to some other unknown phase.
To resolve this issue in the absence of a lattice phase, some additional phase appears to be required, with a different topology altogether. We conjecture that this phase is the lattice we described above, but other candidates could include black holes that are non-uniform but also not localized along the $T^{4}$ directions (see \cite{Dias:2017coo} for a construction in a different context), or a ``black mesh'' configuration (also first hypothesized, to our knowledge, in \cite{Dias:2017coo}).

A second set of concerns pertains to the single black hole phase. It is possible that the assumptions presented in section~\ref{sec:properties_of_new_BH}, as well as the toy model in appendix~\ref{sec:scalar_green_function}, do not capture the correct physics of black holes localized on $T^{4}$. In that case, a single black hole localized on $T^{4}$ could potentially dominate over black holes localized on $S^{3}$, and the naive expectation could be realized.

We hope that the localized black holes discussed in this section can be constructed analytically or numerically, in order to clarify all of these issues, and to prove (or disprove) the phase diagram suggested in figure \ref{fig:microcanonical_ads3_rt_large}.

\section{Conclusions and future directions}
\label{sec:summary}

In this paper we presented the microcanonical phase diagram for the D1-D5 CFT in various regions of the parameter space of this CFT. In the free orbifold limit the phase diagram was analyzed previously in \cite{Hartman:2014oaa}. In the region where supergravity is valid and $R_T \ll R$ the phase diagram is similar to what one may naively expect. When $R_T \gg R$ (and $Q_1 \gg Q_5$ such that the ten-dimensional gravity in this region is under control) we found a large range of energies where a novel phase dominates, described by a lattice (in the $T^4$ directions) of black holes that are localized in $AdS_3$ and in $T^4$ and are uniform on $S^3$. It would be very interesting to analyze in more detail, analytically or numerically, the properties of such black holes, in order to test our conjecture for the phase diagram in this case.

In our paper, we focused on the phases that dominate the microcanonical ensemble at some energy, but we noted that there should also be many additional black hole solutions with different topologies that are never dominant, such as the solutions recently found analytically in \cite{Bah:2022pdn,Bena:2024gmp}. In general, we expect that whenever there is a compact space, at low energies solutions should exist that correspond to black branes wrapping any maximal cycles (not necessarily topologically non-trivial) in the compact space, and that these solutions may continue to exist also at higher energies, until they fill the whole compact manifold; many of the solutions of \cite{Bah:2022pdn,Bena:2024gmp} fall into this class. It would be very interesting if any additional solutions of this type can also be found analytically.

There are various possible generalizations of our analysis. We focused in this paper on the case where the $T^4$ has the same radius for all circles, and it should be interesting to generalize our analysis to cases with different radii (and in particular with some radii larger and some smaller than $R$; all of these configurations are part of the full moduli space of the D1-D5 CFT). It would also be interesting to consider configurations (including black holes) with charge (either under $SU(2)\times SU(2)$, or momentum/winding on the $T^4$) and/or angular momentum, and to figure out the phase diagram as a function of these additional parameters. While we do not expect any of the phases we described in this paper (except the gravitons and BTZ phases) to dominate the canonical ensemble at any temperature, it would be interesting to study the canonical ensemble with chemical potentials for the charge and/or the angular momentum, and to see which solutions dominate there (see \cite{Choi:2024xnv,Bajaj:2024utv} and references therein for a similar analysis for $AdS_5\times S^5$).

The generalization of our analysis to type IIA or type IIB string theory $AdS_3\times S^3\times K3$ backgrounds should be straightforward. A generalization to $AdS_3\times S^3\times S^3\times S^1$ backgrounds, where only the radius of the $S^1$ is a free parameter, but where one of the $S^3$'s can also be much larger than the AdS radius, should also be possible. 

It would be interesting to understand if there are any  $AdS_d\times M$ backgrounds of quantum gravity with $d>3$, in which $M$ can be much larger than the AdS radius, so that similar configurations to the ones we discussed in section \ref{sec:rt_big} are possible. We do not know of any examples.

\acknowledgments
The authors would like to thank Micha Berkooz, Nikolay Bobev, Yacov-Nir Breitstein, Emanuel Katz, Amit Sever and Vit Sriprachyakul for useful discussions, and Oscar Dias and Jorge Santos for useful discussions and comments on a draft of this paper.  
This work was supported in part by ISF grant no. 2159/22, by Simons Foundation grant 994296 (Simons Collaboration on Confinement and QCD Strings), by the Minerva foundation with funding from the Federal German Ministry for Education and Research, and by the German Research Foundation through a German-Israeli Project Cooperation (DIP) grant ``Holography and the Swampland''. OA is the Samuel Sebba Professorial Chair of Pure and Applied Physics.  The work of RF is supported by a research grant from the Chaim Mida Prize in Theoretical Physics at the Weizmann Institute of Science.

\appendix

\section{The scalar spatial Green function on \texorpdfstring{$AdS_{3} \times \mathbb{R}^4$}{ads3 r4}} \label{sec:scalar_green_function}
\subsection{Motivation}

In this section, we present an explicit calculation that motivates the functional structure of the entropy--energy relation in \eqref{eq:new_BH_entropy_energy_general_arguments} for a black hole that is localized in the $AdS_{3}\times T^{4}$ directions and uniform over the $S^{3}$.

Determining exact black hole solutions that are localized along the $T^{d}$ coordinates is a challenging task, both analytically and numerically. Accordingly, we will not attempt to construct such solutions here, and instead leave this as a possible direction for future work.
Rather, we will use heuristic arguments to guess the approximate shape of the horizon of the black holes, and we hope that this captures qualitative features of these black holes and provides useful insight.

In the following, we use the spatial (i.e.\ Poisson) scalar Green function $G$ (with a source at the origin of $AdS_3\times \mathbb{R}^d$) to formulate an approximation for the metric of black holes in $AdS_{3}\times \mathbb{R}^{d}$. The conjectured approximation rests on three assumptions:

\begin{enumerate}
    \item The horizon geometry is approximated by level sets of $G$ (i.e.\ hypersurfaces of constant $G$).
    \item The energy $E$ associated with a given horizon is inversely proportional to the value of $G$ on that horizon. In particular, for a level set defined by $\mu\,G=1$, the corresponding energy scales as $E\propto \mu$.
    \item The horizon area (and hence the entropy $S$) can be approximated by computing the induced area using the vacuum background metric.
\end{enumerate}

These assumptions are motivated by the fact that the linearized equations of motion of the traceless components of the metric are very similar to those of a massless scalar field. In particular, we expect the large-distance dependence of the massless scalar Green function to resemble that of individual metric components, such as $g_{tt}$.

Moreover, one can check that these assumptions are all qualitatively valid both for very small localized black holes in $AdS_3\times \mathbb{R}^d$ (as we show in section \ref{sec:green_function_adm_analog}), and for AdS$_d$-Schwarzschild black holes with $d>3$ at any energy. So, we hope that they will be approximately valid also for the localized black holes that we are interested in, even when they are large (compared to $R$).

However, even if the scalar Green function captures certain features of higher-spin fields at the linearized level, there remains a nontrivial gap between linearized perturbation theory and the full nonlinear equations. Since the horizon is intrinsically a nonlinear phenomenon, it is far from obvious that its area can be reliably estimated using the vacuum background metric, as assumed in the third point -- or, indeed, that the horizon shape is well approximated by the large-distance level sets of $g_{tt}$, as postulated in the first point. Thus we view the approximation we make as a conjecture that remains to be tested more rigorously.

We present the mathematical derivation of the scalar Green function in section~\ref{sec:big_rt_scalar_green_function}, and then apply it, in accordance with the assumptions above, in section~\ref{ref:rt_big_bh_entropy}. In section~\ref{sec:non_optimal_lattice} we compute the entropy of a lattice of black holes that are far enough from each other so that their interactions can be neglected, using the results of section~\ref{ref:rt_big_bh_entropy}.

\subsection{The Green function} \label{sec:big_rt_scalar_green_function}

In this subsection 
we derive an analytic expression for the scalar spatial  Green function on $AdS_3\times \mathbb{R}^d$ and analyze some of its properties. We assume $R_T$ is large and we neglect the effects of the finite torus radius.\footnote{In solving for the Green function, the effects of compactification on a torus of radius $R_{T}$ can be incorporated via the method of mirror charges. These corrections are negligible for our purposes: as we will see, the Green function decays exponentially in the coordinates along the $T^d$ directions, so the mirror charge contributions are suppressed by factors of order $\mathcal{O}\left(\exp(-2\pi R_{T}/R)\right)$.}

\subsubsection{Inversion of the Laplacian} 

We compute the scalar Green function $G$ generated by a \textit{time-independent} source localized at the spatial origin. Accordingly, $G$ is \textbf{not} the scalar propagator; rather, it resembles the static potential produced by a point charge. We assume that the scalar field has mass squared $M^2$, which we set to zero at the end of the computation.

Since we are looking for solutions that are uniform on the $S^3$ and $r \ll R_T$, we ignore the $S^3$ and the periodicity of $T^4$ and solve the Laplace equation on $AdS_3\times \mathbb{R}^d$, where $d=1,\cdots,4$ is the number of directions on $T^4$ that the solution is localized on. We use the metric
\begin{equation} \label{eq: metric_global_ads3_t4}
    ds^{2}=-\left(1+\frac{r^{2}}{R^{2}}\right)dt^{2}+\frac{dr^{2}}{1+\frac{r^{2}}{R^{2}}}+r^{2}d\phi^{2}+\sum_{a=1}^{4}dx_{a}^2,
\end{equation}
where we denote by $x_{a}$ the coordinates on $R^d$ and by $r$ the radial coordinate on $AdS_3$.
The equation we want to solve is\footnote{By $\delta^{\left( 2+d \right)}\left(\vec{x},r\right)$ we mean a source/delta function placed at the origin, normalized such that the integration over the whole space is one. Less formally, we have $\delta^{\left( 2+d \right)}\left(\vec{x},r\right)=\frac{1}{2\pi r}\delta\left(r\right)\delta^{\left(d\right)}\left(\vec{x}\right)$.}
\begin{equation}\label{eq: define G}
    \left(\square_{AdS_{3}\times \mathbb{R}^{d}}-M^{2}\right)G=\mu \delta^{\left(2+d\right)}\left(\vec{x},r\right),
\end{equation}
where $\mu$ is the strength of the source, which we set to $\mu=1$ in this section.
Exploiting the symmetries of the setup, the Green function is independent of $t$ (by construction) and of $\phi$ (by the rotational symmetry of $AdS_3$). Consequently, the $AdS$ contribution to the Laplacian reduces to
\begin{equation}
    \square_{AdS_{3}}G\left(r\right)=\frac{1}{r}\frac{\partial}{\partial r}\left(r\left(1+\frac{r^{2}}{R^{2}}\right)\frac{\partial G\left(r\right)}{\partial r}\right) .
\end{equation}

Performing a Fourier-transform along the $R^d$ directions, \eqref{eq: define G} becomes
\begin{equation}\label{eq: Reduced Klein Gordon}
    \left(\square_{AdS_{3}}-M^2-K^2\right)\tilde{G}=\delta^{\left(2\right)}\left(r\right)
    \ \ ,\ \ G\left( \vec{x},r\right)=\frac{1}{\left(2\pi\right)^d}\int d^d\vec{K}\tilde{G}\left(\vec{K},r\right)e^{iK\cdot \vec{x}}.
\end{equation}
Since $\tilde{G}$ depends only on $K^2$, we can express this Fourier transform as a one-dimensional Hankel transform\footnote{For the case $d=1$, this simplifies to $\frac{1}{\pi R}\int_0^\infty dk\ \tilde{G}\left(k,\rho\right)cos\left(kx\right)$.} 
\begin{equation}\label{eq: Hankel}
G\left(x,\rho\right)=
R^{-d}\frac{1}{x^{\frac{d-2}{2}}\left(2\pi\right)^{\frac{d}{2}}}\int_0^\infty dkk^{\frac{d}{2}}\tilde{G}\left(k,\rho\right)J_{\frac{d-2}{2}}\left(kx\right)
\end{equation}
where we have introduced the dimensionless coordinates\footnote{Note that $x$ is dimensionless, unlike $x_a$ and $x_H$ in the main text.} $\rho=\frac{r}{R}$, $x=\frac{\sqrt{\vec{x}^2}}{R}$ and $k=KR$. 

The solution for $\tilde{G}$ proceeds as follows. Defining $z = 1 + 2\rho^2$ and $m = MR$, and using the dimensionless coordinates introduced above, \eqref{eq: Reduced Klein Gordon} reduces to the Legendre differential equation
\begin{equation} \label{eq: green_fucntion_ledendre_equation}
    \frac{\partial}{\partial z}\left(\left(1-z^{2}\right)\frac{\partial\tilde{G}}{\partial z}\right)+\Delta\left(\Delta+1\right)\tilde{G}=0
\end{equation}
with $\Delta=\frac{1}{2}\left(-1+\sqrt{1+k^{2}+m^{2}}\right)$. 
Equation~\eqref{eq: green_fucntion_ledendre_equation} admits two linearly independent real solutions, and  $\tilde{G}$ can be written as
\begin{equation}
    \tilde{G}
    = c_{1}(\Delta)\, P_{\Delta}(z)
    + c_{2}(\Delta)\, {\rm Re}\!\left[Q_{\Delta}(z)\right] \, .
\end{equation}
Here $P_{\Delta}(z)$ and $Q_{\Delta}(z)$ are the Legendre functions of the first and second kind, respectively, and $c_{1}(\Delta)$ and $c_{2}(\Delta)$ are their corresponding coefficients. Normalizability as $\rho \to \infty$ immediately implies that $c_{1}(\Delta)=0$. To find $c_{2}(\Delta)$, we integrate \eqref{eq: Reduced Klein Gordon} over a disk (in the AdS directions) around the origin with a very small radius $\rho_0 \rightarrow0$. Denoting the region as $B$ and its boundary by $S=\partial B$, we find
\begin{multline}
    1=
    \int_{B}\delta^{\left(2\right)}\left(r\right)=
    \int_{B}\left(\square_{AdS}\tilde{G}-\left(m^{2}+k^{2}\right)\tilde{G}\right)\approx
    \int_{B}\square_{AdS}\tilde{G}=
    \\=
    \int_{S}{\rm grad}\left(\tilde{G}\right) \approx
    \intop_{0}^{2\pi}\rho_{0}\frac{\partial\tilde{G}}{\partial\rho}\left(\rho_0\right)d\phi=
    2\pi\rho_{0}\frac{\partial\tilde{G}}{\partial\rho}\left(\rho_0\right)\approx
    -2\pi\rho_{0}\frac{c_{2}}{\rho_{0}}, 
\end{multline}
which sets $c_{2}(\Delta) = - \frac{1}{2\pi}$.\footnote{As we will see below, the precise value of $c_{2}$ does not enter the computation, since we ultimately normalize the full Green function so that its overall strength matches the energy. What matters is that $c_{2}(\Delta)$ is independent of $\Delta$.}
We end up with
\begin{equation} \label{eq: green_rho_k_end_result}
    \tilde{G} \left(k, \rho \right)=-\frac{1}{2\pi}{\rm Re}\left[Q_{\Delta}\left(1+2\rho^2\right)\right] .
\end{equation}

We find the Green function by integrating \eqref{eq: green_rho_k_end_result} in accordance with \eqref{eq: Reduced Klein Gordon}. A simpler expression can be obtained, which renders the result more practical. To this end, we represent the Legendre functions of the second kind as\footnote{\url{https://functions.wolfram.com/HypergeometricFunctions/LegendreQGeneral/07/01/01/0002/}}
\begin{equation}
Q_{\nu}\left(z\right)=\intop_{0}^{\infty}\left(z+\sqrt{z^{2}-1}\cosh\left(t\right)\right)^{-\nu-1}dt+\frac{1}{2}\left(\log\left(z+1\right)-\log\left(-z-1\right)\right)P_{\nu}\left(z\right) .
\end{equation}
In our case, $z>0$ and $\nu=\Delta>-1$ are real. Hence, this integral representation is valid, and only the first term yields a real contribution. Within this representation, one can perform the integration over $k$ and obtain\footnote{See equation 16 of section 1.4 of \cite{bateman1954tablesI} and equation 19 in section 8.3 of \cite{bateman1954tablesII}. Note that in the latter there is a mistake in the identity, and the exact identity is 
\begin{equation*}
    \intop_{0}^{\infty}y^{1/2}x^{\nu+1}e^{-\alpha\left(x^{2}+\beta^{2}\right)^{1/2}}J_{\nu}\left(xy\right)dx=\left(\frac{2}{\pi}\right)^{1/2}\alpha\beta^{\nu+3/2}y^{\nu+1/2}\left(\alpha^{2}+y^{2}\right)^{-\nu/2-3/4}K_{\nu+3/2}\left(\sqrt{\alpha^{2}+y^{2}}\beta\right),
\end{equation*}
as can be demonstrated by successive application of identities 3.471.15, 6.631.4, and 3.471.9 in \cite{gradshteyn2007}.}
\begin{equation} \label{eq: integral form G}
    G\left(x,\rho\right)=-\frac{1}{2^{\frac{d}{2}}\pi^{\frac{d}{2}+2}}R^{-d}\left(1+m^{2}\right)^{\frac{d}{4}+\frac{1}{4}}\intop_{0}^{\infty}e^{-\beta}\beta\left(\beta^{2}+x^{2}\right)^{-\frac{d+1}{4}}K_{\frac{d+1}{2}}\left(\sqrt{1+m^{2}}\sqrt{\beta^{2}+x^{2}}\right)d\xi\ ,
\end{equation}
with $\beta=\frac{1}{2}\ln\left(1+2\rho^{2}+2\rho\sqrt{\rho^{2}+1}\cosh\left(\xi\right)\right)$.

\begin{figure}[t]
\centering
\includegraphics[width=1\linewidth]{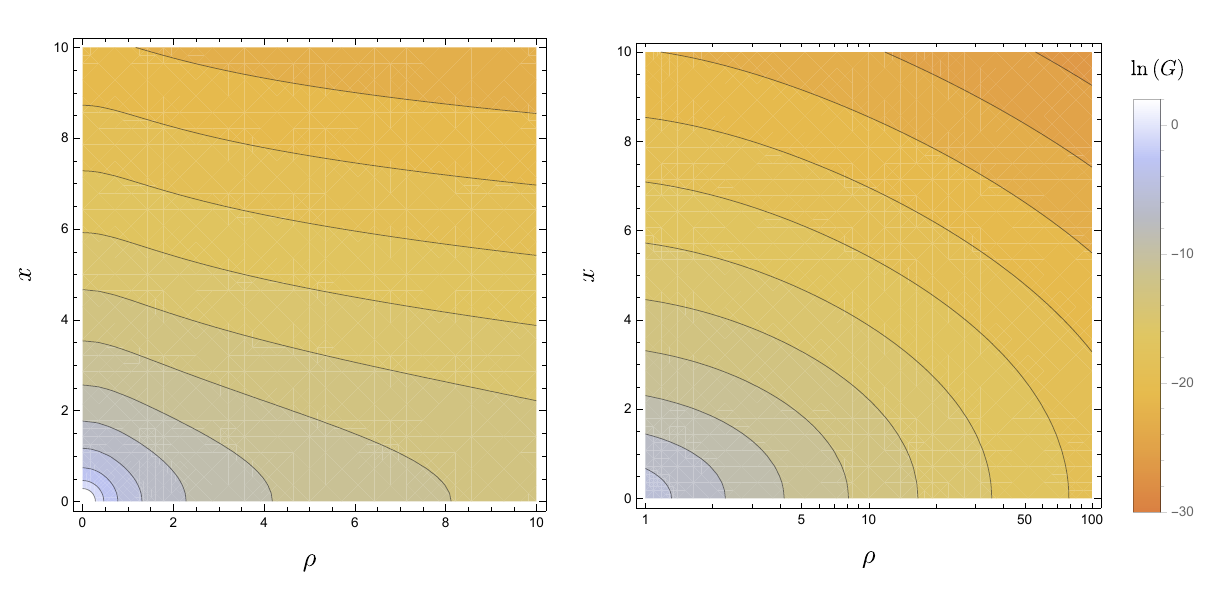}

\caption{\textbf{Left:} Numerically computed curves of the scalar Green function $G\left(x,\rho\right)$ for $d=4$ and $M=0$. As explained in the text, $\rho$ is the radial direction in AdS and $x$ in the $T^4$. For small $x,\rho$ the horizon is spherical, in agreement with the 7d Schwarzschild approximation, while for bigger values the faster growth of the $\rho$ direction compared to $x$ (that was discussed in section \ref{subsec: large x limit}) is apparent. The values in the legends are of $\ln\left(G\right)$. The same qualitative behavior is found for $1\le d<4$. \textbf{Right:} The same plot with log scaling of the $\rho$ axis. 
}
\label{fig:G levels} 
\end{figure}

In the remainder of this subsection, we examine the behavior of the function $G(x,\rho)$ for $m=0$, with particular emphasis on its asymptotic behavior for large $\rho$ and large $x$, as well as on its level curves (contour lines). Representative values of the Green function near the origin are shown in figure~\ref{fig:G levels}, while in figure~\ref{fig:green_function_different_hights} we display the asymptotic shape of the level curves.

\subsubsection{The limit  \texorpdfstring{$\rho\gg1$}{rho gg 1}} \label{subsec: large rho limit}

Here we study the decay of the Green function for $\rho \gg 1$. There are two equivalent ways to obtain this behavior. The first is to take the limit $\rho \to \infty$ directly in momentum space, i.e.\ in \eqref{eq: green_rho_k_end_result}, and only afterwards perform the inverse transform. The second is to take the limit directly in \eqref{eq: integral form G}. Both approaches yield the same result.

Taking $\rho \gg 1$ in \eqref{eq: green_rho_k_end_result}, we obtain (up to an overall constant that depends on $m$)
\begin{equation} \label{eq:green_fucntion_k_large}
    \tilde{G}\left(k,\rho\gg1\right)\approx\frac{1}{2^{2\left(\Delta+1\right)}}\frac{\Gamma\left(-\Delta-\frac{1}{2}\right)}{\Gamma\left(-\Delta\right)}\frac{1}{\tan\left(\pi\Delta\right)}\rho^{-2\left(\Delta+1\right)}.
\end{equation}
Since $\rho$  is large the $k$ integral (recall that $\Delta$ depends on $k$) will rapidly decrease and the major contribution will come from small $k$. In this region, we can expand 
$\Delta\approx-\frac{1}{2}+\frac{1}{2}\sqrt{1+m^{2}}+\frac{k^{2}}{4\sqrt{1+m^{2}}}$ and  $\frac{\Gamma\left(-\Delta-\frac{1}{2}\right)}{\Gamma\left(-\Delta\right)}\approx2\sqrt{\pi}\Delta$ and so \eqref{eq:green_fucntion_k_large} becomes
\begin{equation}
    \tilde{G}\left(k,\rho\gg1\right)\approx e^{-\frac{k^{2}}{2\sqrt{1+m^{2}}}\ln\left(2\rho\right)}.
\end{equation}
The $k$ integral can then be evaluated explicitly and we find that
\begin{equation} \label{eq: large r limit}
G\left(x,\rho\gg1\right)\approx R^{-d}\frac{1}{\rho^{1+\sqrt{1+m^{2}}}\ln^{\frac{d}{2}}\left(2\rho\right)}e^{-\frac{x^{2}\sqrt{1+m^{2}}}{2\ln\left(2\rho\right)}} .
\end{equation}
In the case $x=0$, we find that $G$ decays approximately as a power law. As expected, massive scalar fields exhibit a faster decay than the massless scalar. The integral over $x$ of this expression exhibits the expected behavior for a scalar field of mass $m$ near the boundary of $AdS_3$.

\subsubsection{The limit \texorpdfstring{$x\gg1$}{x gg 1}} \label{subsec: large x limit}

In order to evaluate $G$ in the limit $x \gg1 $ and $\rho \ll 1$, we return to the form \eqref{eq: integral form G}. We expand 
\begin{equation}
    K_{\frac{d}{2}+\frac{1}{2}}\left(\sqrt{1+m^{2}}\sqrt{x^{2}+\beta^{2}}\right)\approx\frac{e^{-\sqrt{1+m^{2}}\sqrt{x^{2}+\beta^{2}}}}{\left(x^{2}+\beta^{2}\right)^{\frac{1}{4}}} ,
\end{equation}
and $\beta\approx\rho \  \cosh\left(\xi\right)$. The $\xi$ integral can then be evaluated explicitly in terms of Appell $F$ functions, but taking again the large $x$ limit we can approximate this as
\begin{equation} \label{eq: large x limit}
    G\left(x\gg1,0\right)\approx R^{-d}\frac{e^{-x\sqrt{1+m^{2}}}}{x^{\frac{d}{2}}}.
\end{equation}

Several features of this limit are noteworthy. First, massive modes again decay faster than massless ones. Second, $G$ decays exponentially in $x$, in contrast to the (almost) power-law behavior in \eqref{eq: large r limit}. This is compatible with \eqref{eq:new_BH_x_as_r_max}, and with our expectations since the theory on $AdS_3$ has an energy gap proportional to $1/R$.

\subsubsection{The limit \texorpdfstring{$x,\rho\ll1$}{x rho ll 1}}

When both $x,\rho\ll1$ we expect the solutions to approach the usual $3+d$-dimensional flat space Green function
\begin{equation} \label{eq:green_function_small_x_
rho}
 G\left(x \ll 1,\rho \ll 1 \right) = \frac{R^{-d}}{\left(x^2+\rho^2\right)^\frac{d}{2}}  ,
\end{equation}
and indeed this is the case, giving a consistency check for the solution \eqref{eq: integral form G}.

\begin{figure}[t]
\centering
\includegraphics[width=0.6\linewidth]{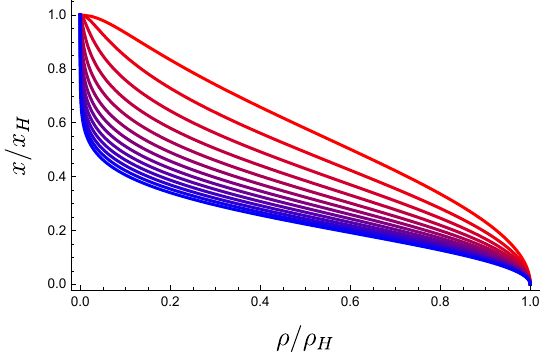}

\caption{
The shape of the level sets for $d=4$ with $M=0$. The axes are normalized so that the maximal values of $\rho$ and $x$ are set to $1$. Curves are shown for $\mu=10^{3}$ (red) through $\mu=10^{15}$ (blue), in multiplicative steps of $10$, with the level chosen as $1/\mu$. As is evident from the figure, as we increase $\mu$ (with the level chosen as $1/\mu$), the shape of the level curves converges, up to overall rescalings of $\rho$ and $x$. The same qualitative behavior is found for $1\le d<4$.
}
\label{fig:green_function_different_hights} 
\end{figure}

\subsection{The entropy-energy relation for black holes in \texorpdfstring{$AdS_{3} \times \mathbb{R}^d $}{ads3 torus}} \label{ref:rt_big_bh_entropy}

Here we use the Green function obtained in the previous section to derive a conjectural relation between the energy and entropy of black holes localized on $T^{4}$. First, in section ~\ref{sec:green_function_adm_analog} we associate an energy to each level set of the Green function. Next, in section \ref{sec:green_function_horizen_area} we compute the horizon area associated with each such level set. Finally, in section \ref{sec:green_function_Entropy_Energy} we combine these results to obtain a conjecture for the relation between energy and entropy.

\subsubsection{The energy from the Green function} \label{sec:green_function_adm_analog}

One can estimate the ADM energy from the Green function $G$. Recall that the energy in AdS$_{d+1}$ space may be read off from the behavior of the $g_{tt}$ component of the metric near the boundary that goes as $1/\rho^d$, since this is the expectation value of $T_{tt}$. In our analogy, we need to look at the behavior of our scalar field near the boundary, and integrate it over the $T^4$ to get the behavior of the lowest KK mode which is the massless scalar field on $AdS_3$.
Because of the Fourier transform formula \eqref{eq: Reduced Klein Gordon} this just amounts to taking $k=0$ (and a factor of $R_T$ for any uniform direction) and $\rho\gg1$. Up to numerical coefficients we then have
\begin{equation}
   \tilde{G}\left(0,\rho\gg1\right)\propto\frac{1}{\rho^{1+\sqrt{1+m^{2}}}},
\end{equation}
which for $m=0$ indeed gives the expected $\frac{1}{\rho^2}$ behavior (on $AdS_3$). This justifies using massless scalars, and implies that the energy associated to a horizon, which we associate with the coefficient of $\frac{1}{\rho^2}$ at $\rho\rightarrow \infty$, should be proportional to the inverse of the value of $G$ at the horizon (the second assumption above).

More concretely, 
as stated in our assumptions, the energy is encoded in the overall normalization $\mu$ of the Green function $\mu G$. For a configuration with a given $\mu(E)$, we expect a horizon to form once the field becomes large, namely when $R^d\mu G$ is of order $\mathcal{O}(1)$. We therefore identify the horizon location with the level set $R^d\mu G=1$, or equivalently $G=R^{-d}/\mu$.
Reinstating the precise factors that go into the ADM energy, we identify the energy of the surface associated to $G=\frac{R^{-d}}{\mu}$ with 
\begin{equation} \label{eq: Energy relation}
    E=\mu\frac{R_{T}^{4-d}R^{3+d}}{G_{10}} ,
\end{equation}
where  
the $R_{T}^{4-d}$ comes from integration over the compact uniform directions, and $R^{3+d}$ was entered to correct the units of $G$.\footnote{We use $R$ and not $R_T$ because, as we defined $G$ in \eqref{eq: define G}, it does not ``see'' the global $R_T$ directions but only the local curvature $R$.}

As a sanity check, note that a small Schwarzschild black hole (with $r_H \ll R$) that is localized in the $AdS_{3}$ directions and in $d$ directions of the $T^{4}$, while being uniform over the $S^{3}$ and over the remaining $4-d$ directions of $T^{4}$, has a horizon radius
\begin{equation}
    r_{H}\propto\left(\frac{G_{10}}{R_{T}^{4-d}R^{3}}E\right)^{\frac{1}{d}}.
\end{equation}
According to our assumptions, the black hole horizon is located on the level set $G = R^{-d}/\mu$. Therefore, for small black holes our toy model predicts a horizon size
\begin{equation}
r_{H} \propto \mu^{1/d}R,
\end{equation}
see \eqref{eq:green_function_small_x_
rho}. Comparing this prediction with the small-Schwarzschild approximation, we obtain the relation
\begin{equation}
    E=\mu\frac{R_{T}^{4-d}R^{3+d}}{G_{10}} ,
\end{equation}
Which is consistent with both \eqref{eq: Energy relation} and \eqref{eq: singlr black hole general form}.

\subsubsection{The horizon area} \label{sec:green_function_horizen_area}
Our next task is to estimate the horizon area as a function of $\mu$. 
As described above, we conjecture that the vacuum metric
\begin{equation}
    ds^{2}=-g_{tt}dt^2+R^2\left(g_{\rho\rho}d\rho^{2}+\rho^{2}d\phi^2+dx^{2}+x^{2}d\Omega_{d-1}\right)
\end{equation}
with $g_{\rho\rho}=\frac{1}{1+\rho^{2}}$ will give a good enough approximation. If we parametrize the horizon by a parameter $a$ such that $\rho(a=0)=\rho_{H}$, $x(a=0)=0$, and $x(a_{\max})=x_{H}$, $\rho(a_{\max})=0$ (i.e.\ $a$ parametrizes the contours and curves shown in figures~\ref{fig:G levels} and \ref{fig:green_function_different_hights}), then the induced metric on the horizon manifold is
\begin{equation}
    \frac{ds_{ind}^{2}}{R^2}=\left[g_{\rho\rho}\left(\frac{\partial\rho}{\partial a}\right)^{2}+\left(\frac{\partial x}{\partial a}\right)^{2}\right]d^{2}a+\rho^{2}d\phi^2+x^{2}d\Omega_{d-1},
\end{equation}
and the area $A$ is
\begin{multline} \label{eq: general horizon area}
    A=
    \int\sqrt{g_{ind}}=
    R^{d+1}\intop_{0}^{2\pi}d\phi\int d\Omega_{d-1}\intop_{0}^{a_{max}}da\ \rho\left(a\right)x\left(a\right)^{d-1}\sqrt{\left(\frac{\partial x}{\partial a}\right)^{2}+g_{\rho\rho}\left(\frac{\partial\rho}{\partial a}\right)^{2}}
    \\=
    2\pi\cdot S_{d-1}\cdot R^{d+1}\intop_{0}^{a_{max}}da\ \rho\left(a\right)x\left(a\right)^{d-1}\sqrt{\left(\frac{\partial x}{\partial a}\right)^{2}+g_{\rho\rho}\left(\frac{\partial\rho}{\partial a}\right)^{2}},
\end{multline}
where $S_{d-1}$ is the area of a unit sphere $S^{d-1}$. We evaluate the area numerically; nevertheless, it is instructive to first estimate the expected functional dependence.

\begin{figure}[t]
\centering
\includegraphics[width=0.45\linewidth]{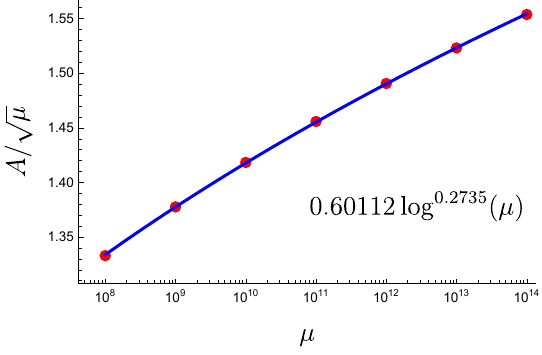}
\includegraphics[width=0.45\linewidth]{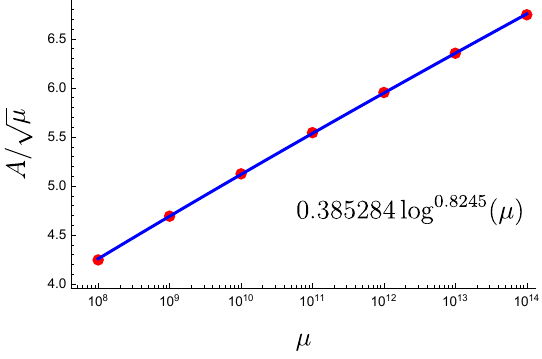}
\includegraphics[width=0.45\linewidth]{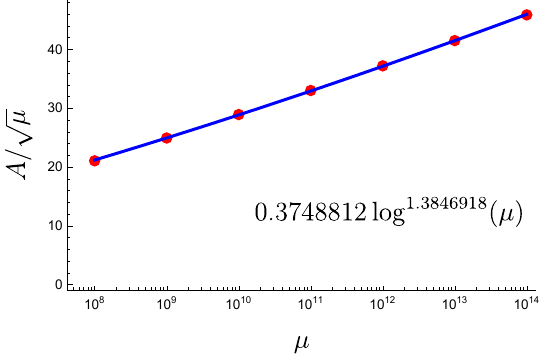}
\includegraphics[width=0.45\linewidth]{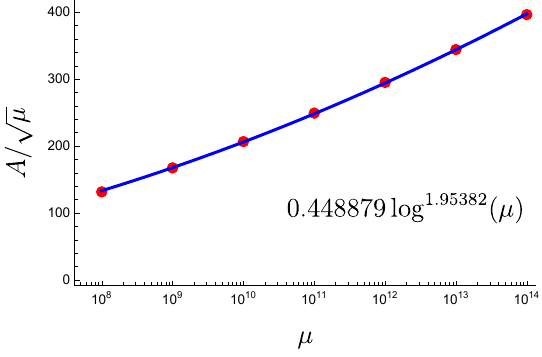}

\caption{Numerical data (red dots) of $\frac{A\left(\mu\right)}{\sqrt{\mu}}$ compared to the fitted power in \eqref{eq:alpha_as_d}
(blue line). \textbf{Top-Left:} $d=1$. \textbf{Top-Right:} $d=2$. \textbf{Bottom-Left:} $d=3$. \textbf{Bottom-Right:} $d=4$
}
\label{fig:fit_A_to_mu} 
\end{figure}

Consider the large-energy regime $\mu\gg 1$. Using the approximations presented in sections~\ref{subsec: large x limit} and \ref{subsec: large rho limit}, and inverting \eqref{eq: large r limit} and \eqref{eq: large x limit}, we find that the maximal values of $\rho$ and $x$ for a given $\mu$ are
\begin{equation} \label{eq:rho_max_and_x_max}
    \rho_{H}\approx\sqrt{\mu}\ln^{-\frac{d}{4}}\left(\mu\right), \quad x_{H}\approx \ln\left(\mu\right) ,
\end{equation}
as was expected on general grounds in \eqref{eq:new_BH_x_as_r_max}.
Applying the triangle inequality $\sqrt{a+b}\leq\sqrt{a}+\sqrt{b}$ to \eqref{eq: general horizon area}, we obtain upper and lower bounds on $A$,
\begin{equation}
    \intop_{0}^{x_{H}}\rho\left(x\right)x{}^{d-1}dx,\intop_{0}^{\rho_{H}}\frac{\rho}{\sqrt{1+\rho^{2}}}x\left(\rho\right)^{d-1}d\rho\lesssim\frac{A}{R^{d+1}}\lesssim\intop_{0}^{x_{H}}\rho\left(x\right)x{}^{d-1}dx+\intop_{0}^{\rho_{H}}\frac{\rho}{\sqrt{1+\rho^{2}}}x\left(\rho\right)^{d-1}d\rho
\end{equation}
Thus, up to numerical prefactors, $\frac{A}{R^{d+1}}\lesssim \rho_{H}x_{H}^{d}$. Using these bounds together with \eqref{eq:rho_max_and_x_max}, we therefore expect that in the regime $\mu\gg 1$ the area approaches the asymptotic behavior
\begin{equation} \label{eq: area_as_mu}
    A\left(\mu\right)\sim R^{d+1} \sqrt{\mu}\ln^{\alpha_d}(\mu)  .
\end{equation}
We verify this numerically by fitting the area to the scaling form \eqref{eq: area_as_mu} over the range $10^{8}<\mu<10^{14}$. This yields a good fit, with the power $\alpha$ given by
\begin{equation} \label{eq:alpha_as_d}
        \alpha_1 \approx 0.27 ,\qquad
        \alpha_2 \approx 0.82 ,\qquad
        \alpha_3 \approx 1.38 ,\qquad
        \alpha_4 \approx 1.95 .
\end{equation}
See figure \ref{fig:fit_A_to_mu} for the explicit fit.

Several comments are in order regarding the results in \eqref{eq:alpha_as_d}. First, concerning the quality of the fit, we find that as we increase the fitting range in $\mu$, the extracted values of $\alpha_{d}$ decrease\footnote{It is possible that the precise form includes also double logarithmic corrections, that do not alter our conclusions.}. Thus, the quoted values should be viewed as upper estimates. Second, as will be important in the next section, all fitted values satisfy $\alpha_{d}<\frac{d}{2}$. Finally, as a consistency check, note that the upper bound on the area contains a factor of $x_{H}^{d}$; therefore, one expects the power of the logarithm to increase with the dimension $d$, in agreement with the trend observed in \eqref{eq:alpha_as_d}.

\subsubsection{The entropy-energy relation}\label{sec:green_function_Entropy_Energy}

We can now combine the results above to obtain an estimate for the entropy. The entropy is given by

\begin{equation}
    S\left(E\right)\sim\frac{R^{3}R_{T}^{4-d}A\left(\mu\left(E\right)\right)}{G_{10}}
\end{equation}
where the factors of $R^{3}R_{T}^{4-d}$ account for the directions along which the configuration is taken to be uniform (a factor of $R^{3}$ from $S^{3}$ and a factor of $R_{T}^{4-d}$ from the remaining $4-d$ directions of $T^{4}$, since we assume the solution is uniform in these directions). Substituting \eqref{eq: Energy relation} in \eqref{eq: area_as_mu}, we obtain that
\begin{equation} \label{eq: entropy_e_localized_end_result}
S\left(E\right)\sim\sqrt{\frac{ER_{T}^{4-d}R^{5+d}}{G_{10}}}\ln^{\alpha_{d}}\left(\frac{EG_{10}}{R^{3+d}R_{T}^{4-d}}\right)\sim c \sqrt{\frac{ER}{c}\left(\frac{R}{R_{T}}\right)^{d}}\ln^{\alpha_{d}}\left(\frac{ER}{c}\left(\frac{R_T}{R}\right)^{d}\right) .
\end{equation}

Equation \eqref{eq: entropy_e_localized_end_result} can raise a concern, as it seems that the entropy of the localized black holes we discuss here grows faster (by logarithmic factors) than the BTZ black hole. This is a potential problem because we know that the BTZ black holes saturates the Cardy formula, which should be a good approximation at energies much larger than $E \sim \frac{c}{R}$ (actually, it is conjectured to be a good approximation even at lower energies \cite{Bena:2024gmp}). On the other hand, if our conjectures hold,  \eqref{eq: entropy_e_localized_end_result} should describe the entropy of the black holes up until $x_{H} \sim \frac{R_T}{R}$, which happens for energies of order $E \sim \frac{c R^{d-1}}{R_T^d} e^\frac{R_T}{R}$. 

Comparing \eqref{eq: entropy_e_localized_end_result} with the BTZ entropy \eqref{eq:ads3_entropy_BH_horizen_s1s3t4} shows that there is, in fact, no inconsistency. In the high-energy regime, the ratio is
\begin{equation}
    \frac{S_{\text{Localized over}\ T^{d}}}{S_{BTZ}}\sim\left(\frac{R}{R_{T}}\right)^{d/2}\log^{\alpha_{d}}(E) .
\end{equation}
Even for extremely large energies $E\sim \exp\left(R_T/R\right)$, the ratio is
\begin{equation}
    \frac{S_{\text{Localized over}\ T^{d}}}{S_{BTZ}}\sim\left(\frac{R}{R_{T}}\right)^{d/2-\alpha_d}.
\end{equation}
Since \eqref{eq:alpha_as_d} implies $\alpha_{d}<\frac{d}{2}$, the BTZ entropy dominates, as expected.

 \subsection{Lattices of black holes with general energy} \label{sec:non_optimal_lattice}

In section \ref{sec:black_hole_lattice} we argued that for energies below $E\sim c/R$ the dominant configuration was a lattice of black holes with individual energy corresponding to some optimal value $\mu_*=\frac{E_*R}{c}\cdot\frac{R_{T}^{4}}{R^{4}}$. At energies $E\sim c/R$ or higher we showed the distance between the black holes in this lattice becomes too small to be able to ignore their interactions. 

What can be done, is to establish a lower bound on the entropy coming from localized black holes at these and higher energies, by computing
the entropy of lattices that contain black holes with individual energies that corresponds to $\mu>\mu_*$, when their spatial spacing is large enough so that we can neglect the interactions between them (this gives a lower bound since the entropy of the dominant localized phase is at least as large as this entropy). The condition for a large enough spatial distance translates into the condition
\begin{equation} \label{eq:mu_constraint_high_energies}
    \mu \gg \frac{ER}{c} \  \ln^{4}(\mu) .
\end{equation}
If we assume the specific form \eqref{eq: entropy_e_localized_end_result}, then 
substituting the constraint \eqref{eq:mu_constraint_high_energies} into \eqref{eq:entropy_of_lattice}, we obtain an upper bound on the entropy of a lattice of black holes with this value of $\mu$,
\begin{equation}
    S \left( E \right) <c\frac{\sqrt{\frac{ER}{c}}}{\ln^{2-\alpha_{4}}\left(\mu\right)} .
\end{equation}
Since $\alpha_{4}<2$, and $\mu$ must grow with the energy to obey \eqref{eq:mu_constraint_high_energies}, we conclude that the entropy of such a phase is consistent with the Cardy bound. 

We may now impose the constraint \eqref{eq:mu_constraint_high_energies} explicitly to obtain a closed-form relation between the entropy and the energy. In particular, the optimal $\mu$ is determined implicitly by
\begin{equation} \label{eq:mu_determination}
    \mu =\tilde{A} \frac{ER}{c} \  \ln^{4}(\mu) ,
\end{equation}
where $\tilde{A}$ is some large number that parametrizes the minimal separation between the black holes that we need in order for our considerations to be reliable. This equation admits an analytic solution in terms of a Lambert W function, and one finds that for large $\mu$ it asymptotes to
\begin{equation}
    \mu\approx 256\tilde{A}\frac{ER}{c}\ln^{\frac{4}{\ln\left(4\sqrt[4]{\tilde{A}\frac{ER}{c}}\right)}+4}\left(4\sqrt[4]{\tilde{A}\frac{ER}{c}}\right) .
\end{equation} 
Consequently, the entropy of a black hole lattice where the black hole separation is characterized by $\tilde{A}$ takes the form (up to double-logarithmic corrections)
\begin{equation}
    \frac{S\left(E\right)}{c} \sim \frac{\sqrt{\frac{ER}{c}}}{8\sqrt{\tilde{A}}\ \ln^{\frac{8}{\ln\left({256\tilde{A}\frac{ER}{c}}\right)}+2-\alpha_{4}}\left(256\tilde{A}\frac{ER}{c}\right)}.
\end{equation}
It seems likely that such a phase never dominates the entropy, but it describes some high-energy states.

\bibliographystyle{JHEP}
\bibliography{biblio.bib}

\end{document}